\documentclass[aps,prx,twocolumn,superscriptaddress]{revtex4-2}
\usepackage{amsmath}
\usepackage{graphicx}
\usepackage{hyperref}
\usepackage{xcolor}
\usepackage{soul}

\newcommand{\bra}[1]{\langle #1 |}
\newcommand{\ket}[1]{| #1 \rangle}
\newcommand{\braket}[2]{\langle #1 | #2 \rangle}
\DeclareMathOperator{\sech}{sech}
\DeclareMathOperator{\csch}{csch}

\begin{document}

\title{Fault-tolerant quantum computation by hybrid qubits \\with bosonic cat-code and single photons}

\author{Jaehak Lee}
\affiliation{Center for Quantum Information, Korea Institute of Science and Technology (KIST), Seoul 02792, Korea}
\affiliation{Department of Physics and Astronomy, Seoul National University, Seoul 08826, Korea}
\author{Nuri Kang}
\affiliation{Center for Quantum Information, Korea Institute of Science and Technology (KIST), Seoul 02792, Korea}
\affiliation{Department of Physics, Korea University, Seoul 02841, Korea}
\author{Seok-Hyung Lee}
\affiliation{Department of Physics and Astronomy, Seoul National University, Seoul 08826, Korea}
\affiliation{Centre for Engineered Quantum Systems, School of Physics, The University of Sydney, Sydney, NSW 2006, Australia}
\author{Hyunseok Jeong}
\affiliation{Department of Physics and Astronomy, Seoul National University, Seoul 08826, Korea}
\author{Liang Jiang}
\affiliation{Pritzker School of Molecular Engineering, The University of Chicago, Chicago 60637 IL, USA}
\author{Seung-Woo Lee}
\email{swleego@gmail.com}
\affiliation{Center for Quantum Information, Korea Institute of Science and Technology (KIST), Seoul 02792, Korea}

\begin{abstract}
Hybridizing different degrees of freedom or physical platforms potentially offers various advantages in building scalable quantum architectures.
We here introduce a fault-tolerant hybrid quantum computation by building on the advantages of both discrete-variable (DV) and continuous-variable (CV) systems. Particularly, we define a CV-DV hybrid qubit with bosonic cat-code and single photon, which is implementable in current photonic platforms. By the cat-code encoded in the CV part, the dominant loss errors are readily correctable without multi-qubit encoding, while the logical basis is inherently orthogonal due to the DV part. We design fault-tolerant architectures by concatenating hybrid qubits and an outer DV quantum error correction code such as topological codes, exploring their potential merits in developing scalable quantum computation. We demonstrate by numerical simulations that our scheme is at least an order of magnitude more resource-efficient over all previous proposals in photonic platforms, allowing us to achieve a record-high loss threshold among existing CV and hybrid approaches. We discuss its realization not only in all-photonic platforms but also in other hybrid platforms including superconducting and trapped-ion systems, which allows us to find various efficient routes towards fault-tolerant quantum computing.
\end{abstract}

\maketitle

\section{\label{sec:Introduction}Introduction}


Towards fully fault-tolerant quantum computing, various physical platforms such as photons \cite{Kok2007,OBrien2009,Slussarenko2019}, superconductors \cite{Krantz2019,Arute2019,Kim2023}, trapped ions \cite{Friis2018,Bruzewicz2019}, nitrogen-vacancies in diamonds \cite{Dutt2007,Doherty2013,Pezzagna2021}, and quantum dots \cite{Petta2005,Chatterjee2021} have been explored and considered as the building block for scalable quantum systems.
Irrespective of the physical platforms, information is encoded into qubits defined with the basis either in discrete-variable (DV) \cite{Browne2005,Dawson2006,Herrera2010,Vigliar2021,Bartolucci2023,Maring2024} or continuous-variable (CV) \cite{Lvovsky2020,Cochrane1999,Jeong2002,Ralph2003,Yokoyama2013,Ulanov2015,Larsen2019,Zhan2020,Larsen2021a,Takeda2019,Enomoto2021,Yonezu2023,Zhong2020,Arrazola2021,Madsen2022} degrees of freedom, each of which has its own pros and cons.
In recent years, hybrid approaches integrating different physical degrees of freedom to overcome the limitations of each platform have opened a new paradigm in quantum technologies. 
Hybridization may be quite a natural direction for scalability, since each platform has its own advantage depending on the circumstances and it is frequently required to convert quantum information between different platforms \cite{Drahi2021,Awschalom2021,Sahu2023}.
Particularly, various CV-DV hybrid protocols have been recently proposed  \cite{Loock2011,Andersen2015,Park2012,LeeSW2013,Kwon2015,Jeong2016,Kim2016,Lau2016,Lim2016,Bergmann2019,Huang2019,Choi2020,Omkar2020,Omkar2021,Wen2021,Bose2022} and experimentally demonstrated \cite{Takeda2013,Jeong2014,Morin2014,Ulanov2017,Jeannic2018,Sychev2018,Cavailles2018,Guccione2020,Gan2020,Darras2023} to combine their advantages in quantum computing and quantum communications.

Meanwhile, qubits encounter errors due to the interaction with environments and imperfect operations, which accumulate and become more severe as increasing the size of the system. Quantum error correction (QEC) provides systematic ways to protect qubits from dominant errors \cite{Nielsen2010,Lidar2013} and allows to achieve fault-tolerance in building scalable quantum architectures. In QEC, information is typically encoded in a Hilbert space larger than a qubit space so that any error can be detected if it brings the encoded state out of the logical code space. By restoring the state back to the code space, errors can be corrected without compromising the encoding of logical information. While multiple physical qubits of finite-dimensional systems are typically used to construct single logical qubit in DV codes, a bosonic system characterized by an infinite-dimensional Hilbert space can provide a large number of degrees of freedom to encode a logical qubit in such a CV approach.
Several bosonic error correction codes, in which a qubit is defined in a single oscillator, have been proposed such as GKP \cite{Gottesman2001}, binomial \cite{Michael2016} and cat code \cite{Cochrane1999,Jeong2002,Ralph2003,Leghtas2013,Mirrahimi2014,Bergmann2016,Li2017}. These codes are elaborately designed, in particular, to protect against photon loss \cite{Albert2018,Terhal2020,Joshi2021}. Recently, bosonic error correction codes have been demonstrated and exploited for building quantum computing systems in various experimental platforms \cite{Leghtas2015,Ofek2016,Lescanne2020,Grimm2020,Fluhmann2019,Campagne2020,Eickbusch2022,Sivak2023,Konno2023, Guillaud2019,Guillaud2021,Chamberland2022,Ni2023,Regent2023,Fukui2017,Fukui2018,Grimsmo2020,Noh2020,Larsen2021b,Bourassa2021,Tzitrin2021,Noh2022}.

In this work, we introduce a hybrid quantum computing scheme by taking the advantages of both CV and DV systems toward fault-tolerant quantum computation. In particular, we define a logical qubit by hybridizing a cat-code encoded state with single photon.
Due to the cat-code encoded in the CV part, the effect of loss is readily correctable even without multi-qubit encoding, while its logical basis is inherently orthogonal due to the DV part in contrast to other CV qubits \cite{Cochrane1999,Jeong2002,Ralph2003,Ourjoumtsev2007,Nielsen2010,Leghtas2013,Mirrahimi2014,Bergmann2016,Li2017}. Such hybrid qubits can be experimentally generated in current photonic platforms  \cite{Jeong2014,Morin2014,Guccione2020,Darras2023}.
We then design quantum computing architectures by concatenating the hybrid qubits and outer DV quantum error correction code such as Steane code \cite{Steane1996} and surface code \cite{Raussendorf2006,Raussendorf2007,Fowler2009}.
Specifically, we here focus on all-photonic implementations for direct comparison with previous proposals \cite{Dawson2006,Herrera2010,Lund2008,LeeSW2013,Omkar2020}. To that end, we define hybrid fusion schemes as building blocks for scalable architectures, which are implementable in current experimental platforms with linear optics \cite{Jeong2014,Morin2014}. 
We then numerically simulate and analyze the fault-tolerance of our hybrid scheme, showing that it is at least an order of magnitude more resource-efficient over all previous proposals in photonic platforms, owing to the use of the bosonic cat-code in physical level. Moreover, it is demonstrated that our scheme allows to achieve at least 4-times higher loss thresholds compared to existing hybrid and CV approaches ~\cite{Lund2008,LeeSW2013,Omkar2020}. 
We stress that our scheme is not limited to all-photonic platforms but can be implementable in other hybrid platforms including superconducting \cite{Leghtas2015,Ofek2016,Lescanne2020,Grimm2020,Vlastakis2013,Wang2016,Wang2022,He2023,Pan2023} and trapped-ion systems \cite{Gan2020,Eickbusch2022,Monroe2021,Heeres2017,Blais2021,Monroe1996,Kienzler2016,Johnson2017,Jeon2024}, which thus allows us to find various routes toward scalable fault-tolerant quantum computing.

\section{\label{sec:Hybrid}Hybrid qubits for quantum computation}

\subsection{\label{sec:hyQubit}Hybrid qubit}

\begin{figure}[t] 
        \centering \includegraphics[clip=true, width=3.2cm]{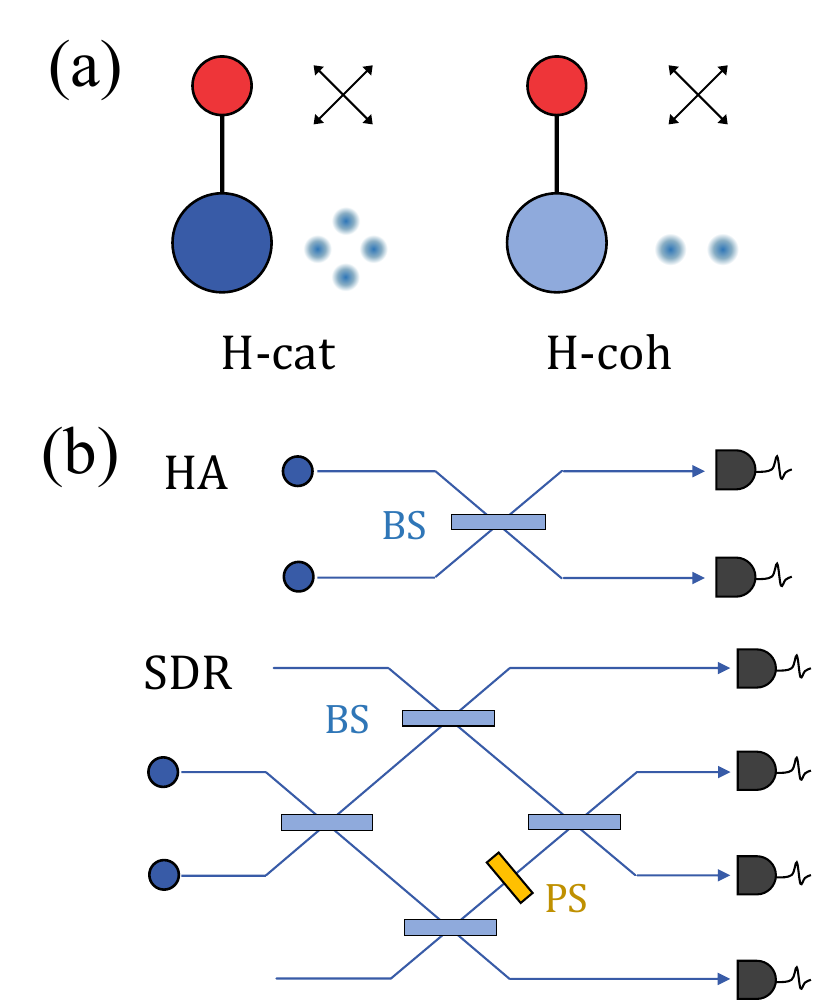}
        \hspace{0.1cm} \includegraphics[clip=true, width=5.1cm]{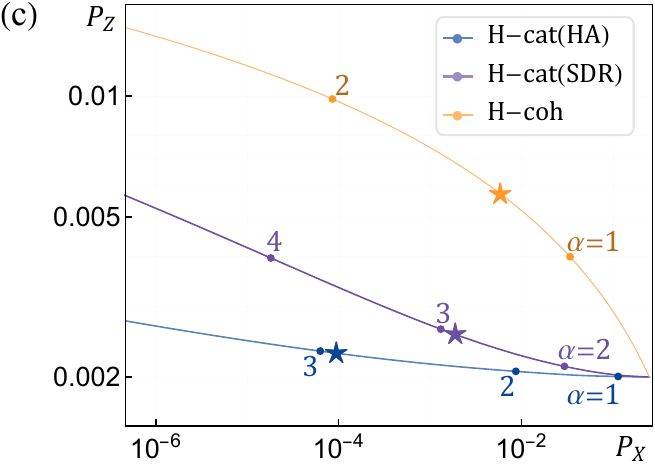}
        \caption{(a) An illustration of hybrid qubit. In our H-cat qubit, DV(red) and CV(blue) qubits are encoded, respectively, in the polarization degree of freedom and in the even cat states $\{\ket{\mathcal{C}_{\alpha}^+},\ket{\mathcal{C}_{i\alpha}^+}\}$, while CV qubit is encoded in the coherent states(light blue) in the H-coh qubit. (b) Schemes for fusion operation of CV qubits. We employ two separate schemes of cat-code Bell measurement which we respectively refer to HA scheme \cite{Hastrup2022} and SDR scheme \cite{Su2022}. BS and PS represent beamsplitter and $\pi/2$-phase shifter, respectively. (c) The plot represents the error rates $P_X$ and $P_Z$ for the fusion of hybrid qubits under loss rate $ \eta = 2\times10^{-3} $. The orange curve is for H-coh scheme and the blue(purple) curve is for H-cat scheme employing HA (SDR) scheme. Numbers indicated at each point represent the corresponding amplitude $\alpha$. Points marked as stars represent the optimal encoding amplitude $\alpha$ which achieves the highest loss threshold.}
        \label{fig:Hybrid}
    \end{figure}

We define a hybrid qubit by employing single photon (DV qubit) and cat-code encoded state (CV qubit), which we call the hybrid cat-code (H-cat) qubit. 
The cat-code is a bosonic QEC code designed to protect qubits against photon loss, in which qubits are encoded in the subspace of even photon number states so that a photon loss can be detected when the parity changes \cite{Leghtas2013,Mirrahimi2014}. Specifically, the code space is defined by even cat states $ \left\{ \ket{\mathcal{C}_{\alpha}^+} , \ket{\mathcal{C}_{i\alpha}^+} \right\} $, where $ \ket{\mathcal{C}_{\alpha}^+} = \mathcal{N}_\mathcal{C}^+ (\ket{\alpha}+\ket{-\alpha}) $ with the normalization factor $ \mathcal{N}_\mathcal{C}^+ \equiv 1/\sqrt{2(1+e^{-2\alpha^2})} $ and $ \ket{\alpha} $ is a coherent state with amplitude $ \alpha $. 
By combining DV and CV qubits, the basis of H-cat qubit can then be defined as
    \begin{equation}
    \label{eq:hyqubit}
        \left\{ \ket{0_L} = \ket{+}\ket{\mathcal{C}_{\alpha}^+} , \ket{1_L} = \ket{-}\ket{\mathcal{C}_{i\alpha}^+} \right\},
    \end{equation}
where $ \ket{\pm} = ( \ket{H} \pm \ket{V} )/\sqrt{2}$ is the polarization of single photon. Note that the logical basis are orthogonal to each other thanks to the DV part in contrast to the basis $\{\ket{\mathcal{C}_{\alpha}^+}, \ket{\mathcal{C}_{i\alpha}^+}\}$ in bosonic quantum computation which overlaps each other with finite $\alpha$ and causes inherent errors.
We also consider another type of hybrid qubits \cite{LeeSW2013} for comparison, composed of single photon and coherent state in the basis $\{\ket{+}\ket{\alpha} , \ket{-}\ket{-\alpha}\}$, which we call here the hybrid coherent-state (H-coh) qubit. The hybrid qubits, i.e.,~H-cat and H-coh, are illustrated in Fig.~\ref{fig:Hybrid}(a).


Note that hybrid qubits can be generated in current photonic platforms. CV-DV hybrid entangled states have been experimentally demonstrated in optical systems  \cite{Jeong2014,Morin2014} and successfully applied to quantum computing and communications in numerous experiments~\cite{Huang2019,Guccione2020,Darras2023}. In those experiments, the CV basis is defined by coherent states i.e.~$\ket{\alpha}$ and $\ket{-\alpha}$, constituting two-component cat states in CV part, so that the generated hybrid entangled states can be directly used as the H-coh qubit by simple modifications.
We can also generate the H-cat qubit by using H-coh qubits by extending two-component cat states into four-component using $\ket{\alpha}$, $\ket{i\alpha}$, $\ket{-\alpha}$, and $\ket{-i\alpha}$ in CV part, e.g.,~using two H-coh qubits, a beam splitter and a photon number resolving detector as proposed in Ref.~\cite{Hastrup2020}. We present the scheme for H-cat qubit generation in Sec.~\ref{sec:Resource}.

We stress that such CV-DV hybrid entanglement can be generated efficiently also in other platforms including superconducting and trapped-ion systems, which enables that our approach can be more generally implemented in any CV-DV hybrid platforms for quantum computing. We discuss these further in Sec.~\ref{sec:Implementation}.

\subsection{\label{sec:hyFusion}Hybrid fusion}

We now introduce a \emph{hybrid fusion}, i.e.,~a CV-DV hybrid entangling operation, which is performed  by a joint work of Bell-state measurements on CV and DV qubits. A fusion measurement is typically applied on entangled states to generate larger size entangled states such as cluster states as prerequisites for measurement-based quantum computation (MBQC) \cite{Raussendorf2001,Menicucci2006,Larsen2021b}, or also enables universal gate operations via teleportation in circuit-based quantum computation \cite{Gottesman1999}.

A hybrid fusion projects a two hybrid qubits on one of the hybrid Bell states $\ket{\Psi_L^\pm} =( \ket{0_L}\ket{1_L} \pm \ket{1_L}\ket{0_L} )/\sqrt{2}$, $\ket{\Phi_L^\pm} =( \ket{0_L}\ket{0_L} \pm \ket{1_L}\ket{1_L} )/\sqrt{2}$. The logical Bell states of hybrid qubits given in the basis of Eq.~(\ref{eq:hyqubit}) can be rewritten by separating the CV and DV parts as (details are in the Appendix~\ref{sec:FusionHybrid})
    \begin{equation}
    \label{eq:hyBellStates}
    \begin{aligned}
        \ket{\Psi_L^\pm} & \propto \ket{\Phi_\mathcal{D}^-}\ket{\tilde{\Psi}_\mathcal{C}^\pm} + \ket{\Psi_\mathcal{D}^-}\ket{\tilde{\Psi}_\mathcal{C}^\mp} \\
        \ket{\Phi_L^\pm} & \propto \ket{\Phi_\mathcal{D}^+}\ket{\tilde{\Phi}_\mathcal{C}^\pm} + \ket{\Psi_\mathcal{D}^+}\ket{\tilde{\Phi}_\mathcal{C}^\mp} .
    \end{aligned}
    \end{equation}
where $ \ket{\Psi_\mathcal{D}^\pm} = ( \ket{H}\ket{V} \pm \ket{V}\ket{H} )/\sqrt{2}$, $ \ket{\Phi_\mathcal{D}^\pm} = ( \ket{H}\ket{H} \pm \ket{V}\ket{V} )/\sqrt{2} $ are the Bell states of DV qubits, and $ \ket{\tilde{\Psi}_\mathcal{C}^\pm} = \ket{\mathcal{C}_{\alpha}^+}\ket{\mathcal{C}_{i\alpha}^+} \pm \ket{\mathcal{C}_{i\alpha}^+}\ket{\mathcal{C}_{\alpha}^+} $, $ \ket{\tilde{\Phi}_\mathcal{C}^\pm} = \ket{\mathcal{C}_{\alpha}^+}\ket{\mathcal{C}_{\alpha}^+} \pm \ket{\mathcal{C}_{i\alpha}^+}\ket{\mathcal{C}_{i\alpha}^+} $ are unnormalized Bell states of CV qubits. 

Therefore, the hybrid fusion can be implemented by applying CV and DV Bell-state measurements separately, denoted here as $\mathrm{B}_\mathcal{C}$ and $\mathrm{B}_\mathcal{D}$, respectively. Its logical outcome can then be discriminated by combining the results of $\mathrm{B}_\mathcal{C}$ and $\mathrm{B}_\mathcal{D}$ based on Eq.~(\ref{eq:hyBellStates}). 
$\mathrm{B}_\mathcal{C}$ can be implemented by linear optics and photon-number-resolving (PNR) detectors as illustrated in Fig.~\ref{fig:Hybrid}(b). Two schemes were recently proposed independently in Ref.~\cite{Hastrup2022} and Ref.~\cite{Su2022}, we refer to them as the HA and SDR scheme, respectively (for details, see Appendix~\ref{sec:FusionCV}).
$\mathrm{B}_\mathcal{C}$ can discriminate the Bell states through the pattern of the measurement outcomes as denoted in the tables of Fig.~\ref{fig:Fusion}(b,c); while the logical sign $ +/- $ of Bell states can be distinguished with certainty by counting the total number of photons detected, there exists ambiguity in discriminating the logical letter $ \Psi/\Phi $ for some measurement outcomes for which we assign the $X$ error with the rate $p_X$.

In our hybrid scheme, we can remove the ambiguity by $\mathrm{B}_\mathcal{D}$. In the photon polarization encoding, $\mathrm{B}_\mathcal{D}$ can be chosen as a so called type II fusion \cite{Browne2005} that distinguishes two Bell states $ \ket{\Psi_\mathcal{D}^\pm} $ out of four with linear optics \cite{Lut99,Calsamiglia2001} (see Appendix~\ref{sec:FusionDV}). Once $\mathrm{B}_\mathcal{D}$ succeeds, it removes the ambiguity between the letter $ \Psi/\Phi $. 
Remarkably, a hybrid fusion is thus able to distinguish hybrid Bell states with certainty even if only one of $\mathrm{B}_\mathcal{C}$ and $\mathrm{B}_\mathcal{D}$ succeeds. Otherwise, i.e., in the case that $\mathrm{B}_\mathcal{C}$ yields ambiguity and $\mathrm{B}_\mathcal{D}$ fails, the resulting state after the hybrid fusion can be represented by $ \mu\ket{\Phi_L^\pm} + \nu\ket{\Psi_L^\pm} $. So, when $ |\mu| \geq |\nu| $ we can take $ \ket{\Phi_L^\pm} $ as the outcome and assign $X$ error with rate $ P_X = |\nu|^2/(|\mu|^2+|\nu|^2) $, and vice versa. We calculate the $X$ error rate of hybrid fusion employing HA or SDR scheme as described in Appendix~\ref{sec:Fusion}.

Moreover, in the presence of photon loss, $\mathrm{B}_\mathcal{C}$ can detect a single photon loss in CV part through detecting the parity change of the cat state to odd, i.e., when a total odd number of photons is detected. This is thanks to the bosonic cat-code error correction in the CV part. 
The hybrid fusion thus can yield the outcome successfully that contains $X$ error with a slightly changed rate from $ P_X $ (see Appendix~\ref{sec:FusionCV}). If two or more photons are lost, $Z$ errors remain undetected in the fusion measurement outcome, as we analyze in the following section.

\subsection{\label{sec:hyError}Error analysis}

If more than one photons are lost, errors can be modeled by applying higher power of $ \hat{a} $. In CV part, the cat-code exhibits a cyclic behavior \cite{Bergmann2016}, that is, the state after $ 4k+l $ photon loss has the same form as the state after $l$ photon loss, where $k$ is a nonnegative integer and $l=0,1,2,3$. Since each photon loss induces a parity change, the CV qubit is in even(odd) space when $l$ is even(odd). $Z$ error may occur by multiple photon loss as $\mathrm{B}_\mathcal{C}$ only corrects single photon loss detected by the parity change. If two photons are lost ($l=2$) at one input port, a $Z$ error occurs as $ \hat{a}^2( a\ket{\mathcal{C}_{\alpha}^+} + b\ket{\mathcal{C}_{i\alpha}^+} ) = ( a\ket{\mathcal{C}_{\alpha}^+} - b\ket{\mathcal{C}_{i\alpha}^+} ) $ and it remains undetected because there is no parity change. When one photon is lost from each CV input port, the total photon number is even, but the parity change may be distinguishable by some measurement patterns (see Appendix~\ref{sec:FusionCV}), which causes no error. If the parity change is not distinguishable, a $Z$ error remains undetected. If $l=3$, only single photon loss is corrected and thus a $Z$ error remains uncorrected as in the case of $l=2$. For the case of higher photon loss, we can make a similar error analysis, and we summarize it in Appendix~\ref{sec:ErrorModel}.
On the other hand, if any of DV photon is lost, we lose the phase information, which causes a dephasing error. DV photon loss is locatable by counting the total number of photons detected in $\mathrm{B}_\mathcal{D}$. To sum up, we denote $P_Z$ to be the overall $Z$ error rate induced by photon loss. We derive the explicit expression of the $X$ error rate $P_X$ and the $Z$ error rate $P_Z$ in Appendix~\ref{sec:ErrorModel}.

In Fig.~\ref{fig:Hybrid}(c), we present the $X$ and $Z$ error rates under loss for the hybrid fusions. Specifically, $P_X$ and $P_Z$ are plotted for the fusions of H-cat qubits with HA and SDR, and H-coh qubit used in Ref.~\cite{LeeSW2013,Omkar2020,Omkar2021} by varying the amplitude $\alpha$ under a fixed loss rate $ \eta=2\times10^{-3} $. 
It shows that the effect of loss can be substantially suppressed in the hybrid fusion of H-cat compared to H-coh thanks to the bosonic cat-code error correction in the CV part. As common tendencies, $P_X$ error can be exponentially suppressed as $\alpha$ grows because the basis of CV part are more distinguishable, while $P_Z$ only increases linearly with $ |\alpha|^2 $ because its state becomes more fragile against photon loss. The result clearly shows that $P_Z$ is suppressed by employing the cat-code error correction. By taking larger encoding amplitude $\alpha$, we can always achieve much smaller $P_X$ and $P_Z$ in the fusions of H-cat qubits than H-coh qubits. 
By comparing two different types of the fusion schemes of H-cat qubits, i.e.~HA and SDR, HA scheme exhibits lower error rates, whereas we show that the SDR scheme is advantageous when an unambiguous discrimination is required without error (see Appendix~\ref{sec:FusionCV}).

\section{\label{sec:FTQC}Fault-tolerant quantum computation}

\subsection{\label{sec:Architecture}Hybrid quantum computing architectures}

Let us now design quantum computing architectures based on the hybrid qubits. We can consider both the circuit-based and MBQC models. 
In the circuit-based model, the gate operations for universal quantum computation can be chosen as the gate set $ \{ \hat{X}, \hat{Z}_\theta, \hat{H}, \hat{CZ} \} $. The Pauli $ X $ operation ($\hat{X}$) and arbitrary rotation along $ Z $ axis ($\hat{Z}_\theta$) are implementable only using linear optical elements; $\hat{X}$ can be implemented by applying bit flip operations in both CV and DV parts, using a polarization rotator acting as $ \ket{+} \leftrightarrow \ket{-} $ and a $ \pi/2 $ phase shifter acting as $ \ket{\mathcal{C}_{\alpha}^+} \leftrightarrow \ket{\mathcal{C}_{i\alpha}^+} $. $ \hat{Z}_\theta $ can be implemented by a $ \theta $ phase shifter applied only as $ \ket{-} \to e^{i\theta}\ket{-} $ on the DV part. The Hadamard ($ H $) and Controlled-$Z$ ($ CZ $) gates can be performed based on the gate teleportation technique \cite{Gottesman1999}, using the hybrid fusion scheme and entangled resource states. The resource states as the channel of teleportation for $ H $ and $ CZ $ are given respectively by $ \ket{\Phi_H} \propto \ket{0_L,0_L} + \ket{0_L,1_L} + \ket{1_L,0_L} - \ket{1_L,1_L} $ and $ \ket{\Phi_{CZ}} \propto \ket{0_L,0_L,0_L,0_L} + \ket{0_L,0_L,1_L,1_L} + \ket{1_L,1_L,0_L,0_L} - \ket{1_L,1_L,1_L,1_L} $. We present the scheme for resource state generation in the next section. The measurement on the $Z$ basis can be performed by the measurement in DV part using polarization measurement. See Appendix~\ref{sec:GateOp} for the details of the schemes of hybrid teleportation and measurement.

Constructing a fault-tolerant architecture in our hybrid approach requires to concatenate the hybrid qubits incorporating the bosonic cat-code and an outer DV quantum error correction code such as CSS \cite{Steane1996} and topological codes \cite{Raussendorf2006,Raussendorf2007,Fowler2009}. For a direct comparison with previous works~\cite{Dawson2006,Herrera2010,Lund2008,LeeSW2013,Omkar2020}, we construct a circuit-based fault-tolerant model using the 7-qubit Steane code (see Appendix~\ref{sec:Simulation}), while we employ the surface code for MBQC as described below. 

In MBQC, we construct a RHG lattice embedding the surface code as shown in Fig.~\ref{fig:RHG}(a). We first prepare 3-qubit micro cluster states by using hybrid qubits as element resources, as we describe in Sec.~\ref{sec:Resource}, which we refer to as 3-qubit micro {\em H-cluster} states. 
We then merge such micro H-cluster states using hybrid fusion to construct a RHG lattice: In step 1, three micro H-cluster states are merged by two hybrid fusions to form a 5-qubit star H-cluster state with one central qubit and four side qubits. In step 2, we pile star H-cluster states in the form of RHG lattice, where the central hybrid qubits become vertices of cluster states and the side qubits are consumed by hybrid fusion to create edges between adjacent vertices.
We note that the deterministic nature of the hybrid fusion enables a resource-efficient construction of the RHG lattice. Moreover, photon loss can be corrected to some extent by implementing the hybrid fusion so that higher loss thresholds can be reached which we will discuss in the following subsection.

In the RHG lattice, each $xy$-plane is called a layer and $t$-axis represents the simulated time. The size of RHG lattice is characterized by the code distance $d$. A layer is primal (dual) if it is located at even (odd) $t$ and primal (dual) qubits reside on faces (edges) of the primal lattice.
The stabilizer of a primal unit cell is given by the product of six Pauli $X$'s on primal qubits located on the faces. The error patterns matching the stabilizer measurement can be corrected and the remaining error chain connecting two primal boundaries causes a logical error. Note that we here only simulate $Z$ errors in primal lattices, because error corrections are done separately on the primal and dual lattice and qubits are measured in $X$ basis for stabilizer measurements.

\subsection{\label{sec:FaultTolerance}Fault-tolerance analysis}


    \begin{figure*}[t] 
        \centering \includegraphics[clip=true, width=10cm]{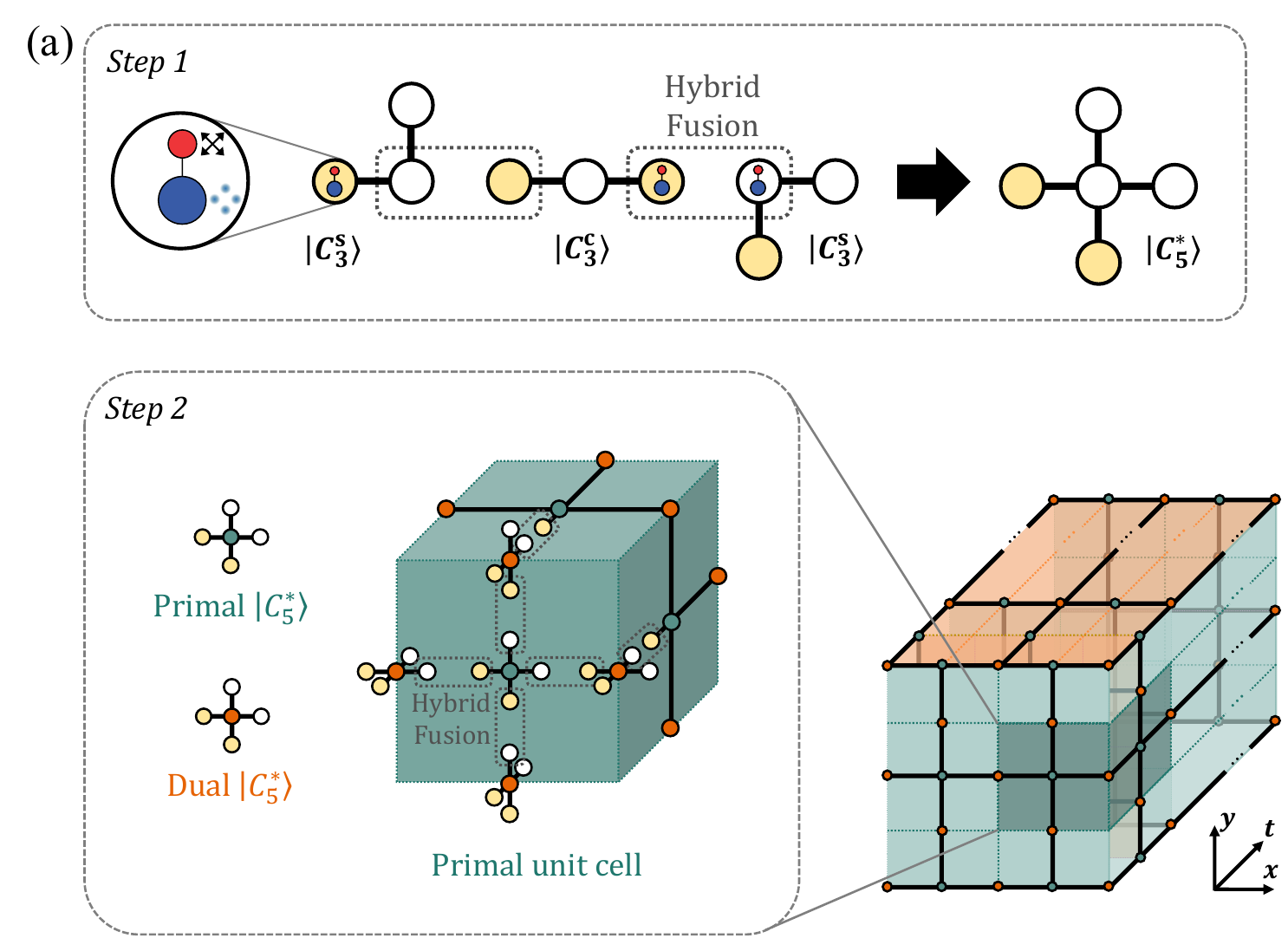}
        \hspace{0.2cm} \includegraphics[clip=true, width=6.5cm]{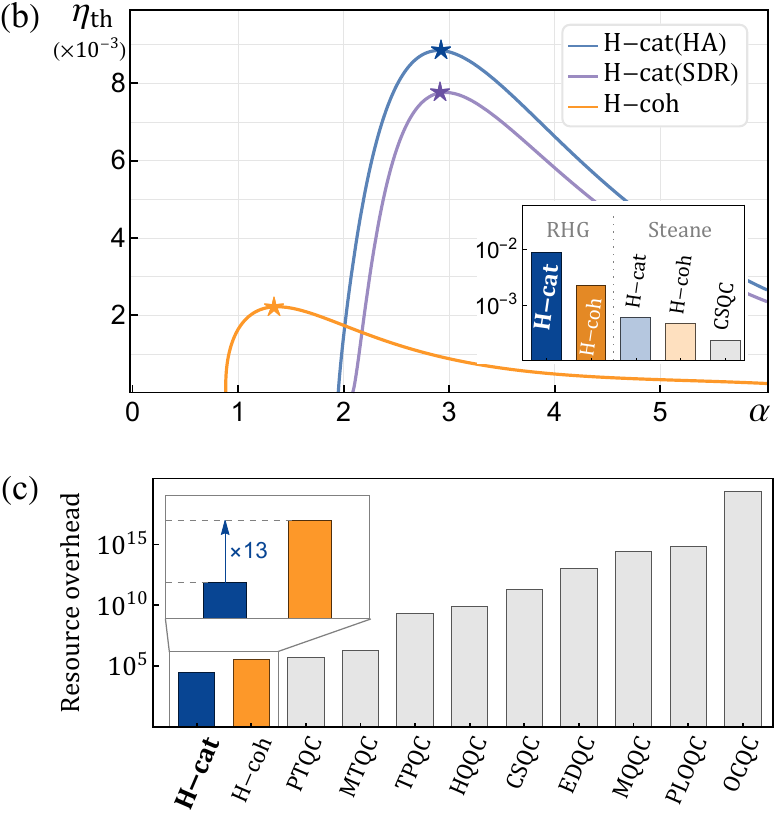}
        \caption{(a) A schematic to build a RHG lattice by hybrid qubits. Step 1: one central micro H-cluster state $\ket{C_3^\mathrm{c}}$ and two side micro H-cluster states $\ket{C_3^\mathrm{s}}$ are merged using two hybrid fusions to create a star H-cluster state $\ket{C_5^\ast}$. Micro cluster states differ by Hadamard gates, which are applied on qubits filled with yellow, from typical cluster states. Step 2: in a unit cell of RHG lattice, qubits are located at each face and edge. Star H-cluster states are placed on each location and merged by hybrid fusion on side qubits. (b) Loss thresholds obtained by simulation in RHG lattice with different schemes. Points marked as stars represent the highest loss threshold. The inset represents the comparison of optimal thresholds including previous approaches with Steane code. (c) Comparison of the resource overheads of existing photonic quantum computing proposals to achieve the logical error rate $ p_L = 10^{-6} $. Hybrid quantum computing with H-cat qubits requires 13 times less resource compared to H-coh qubits when H-coh pair is chosen to be the common resource state. We also compare with other existing photonic schemes: PTQC\cite{LeeSH2023}, MTQC\cite{Omkar2022}, TPQC\cite{Herrera2010}, HQQC\cite{LeeSW2013}, CSQC\cite{Lund2008}, EDQC\cite{Cho2007}, MQQC\cite{LeeSW2015}, PLOQC\cite{Hayes2010}, and OCQC\cite{Dawson2006}.}
        \label{fig:RHG}
    \end{figure*}


We investigate the fault-tolerance of quantum computation with H-cat and H-coh qubits and compare the result with previous works. We here focus on analyzing and simulating the fault-tolerance of MBQC, which is suitable for photonic platforms. For the circuit-based model, we can employ the telecorrection \cite{Knill2005} and Steane code in simulation equivalently with the works using H-coh and CV qubits~\cite{Lund2008,LeeSW2013} for the direct comparison, details of which are in Appendix~\ref{sec:Simulation}. 

In the RHG lattice, errors that occur in the hybrid fusion propagate to adjacent qubits (See Appendix~\ref{sec:ErrorFusion}), so that a corresponding error rate is assigned to each individual qubit on the lattice. Based on assigned error rates, we can find the error pattern matching the syndrome measurement using the weighted minimum-weight perfect matching \cite{Fowler2012} in the PyMatching package \cite{Higgott2021}, and then count remaining error chains connecting two primal boundaries and determine whether a logical error occurs. We perform a Monte Carlo simulation to find the logical error rate $p_L$ for different code distances $d$, and then investigate whether the errors are accumulated, i.e.,~$p_L$ increases or not as increasing $d$. The fault-tolerance noise thresholds can then be determined as the maximum physical error rates by which the logical errors are not accumulated with $d$. See Appendix~\ref{sec:Simulation} for details.

We present the loss thresholds of photonic MBQC based on H-cat and H-coh qubits in Fig.~\ref{fig:RHG}(b) by changing the encoding amplitude $\alpha$ in CV part.
Note that the previous estimation of H-coh in Ref.~\cite{Omkar2020} employed a slightly different error model, and thus we resimulate the result under the same error model assigned to H-cat for fair comparison. We also depict the loss thresholds of the circuit-based model with hybrid and CV qubits estimated based on Steane codes for comparison.

It shows that hybrid MBQC with H-cat qubits achieves the highest loss thresholds over other approaches including MBQC with H-coh qubits as well as all hybrid and CV approach in circuit-based model. The loss threshold of quantum computing with H-cat 0.89\% is about 4-times larger than the maximum 0.22\% estimated with H-coh \cite{Omkar2020}. The hybrid fusion with HA scheme yields higher threshold than SDR scheme as the former exhibits lower error rates than the latter.

The highest thresholds can be reached by an optimized encoding of $\alpha$ marked by stars in Fig.~\ref{fig:RHG}(b) and in Fig.~\ref{fig:Hybrid}(c); $\alpha \approx 2.93$ for H-cat with HA scheme and $\alpha \approx 1.37$ for H-coh qubits. The threshold curves of H-cat and H-coh show a common tendency that they rapidly increase up to the peak and decrease gradually. This is because enlarging the encoding amplitude $\alpha$ can reduce $P_X$ but enhance $P_Z$ so that further increase of $\alpha$ at some point degrades the thresholds eventually. As exhibited in Fig.~\ref{fig:Hybrid}(c), the enhancement of loss thresholds can be achieved with H-cat qubits as $P_X$ can be significantly reduced further while suppressing $P_Z$ by increasing $\alpha$ thanks to the inherent bosonic error correction in CV part.

\subsection{\label{sec:Overhead}Resource overhead}

To investigate the resource overhead, we estimate the number of unit resources referred to as $\mathcal{N}_{p_L}$ to achieves the target logical error rate $p_L$. We choose H-coh pair as a unit resource for fair comparison with the resource estimation in previous works \cite{LeeSW2013,Omkar2020}. 
We propose a scheme to generate an H-cat pair consuming 8 H-coh pairs as illustrated in Fig.~\ref{fig:Resource}(a), details of which are also explained in Sec.~\ref{sec:Resource}. Two H-coh pairs can be merged by applying $\mathrm{B}_\mathrm{I}$ on DV and beamsplitter interaction followed by photon number counting on CV qubits. The resulting state can be postselected with the probability approximately 1/4 to obtain an H-cat pair. Thus we here estimate that an H-cat pair is 8 times as expensive as an H-coh pair.

We find that a target logical error rate can be achieved even with very small code distance $d$ with H-cat qubits. For example, to achieve $ p_L=10^{-6} $, only $d=4$ is sufficient with H-cat qubits due to the suppressed errors by the bosonic cat-code in CV part of the physical level, while $d=15$ is required with H-coh qubits \cite{Omkar2020}. As a result, we can estimate that the resource overhead for the hybrid MBQC with H-cat qubits is $ \mathcal{N}_{10^{-6}}=2.7\times10^4 $,  which is 13 times more efficient compared to the overhead for the MBQC with H-coh qubits $ \mathcal{N}_{10^{-6}}=3.6\times10^5 $. Remarkably, employing H-cat qubits can reduce an order of magnitude resource cost compared to the approach with H-coh qubits.

In Fig.~\ref{fig:RHG}(c), we present the resource cost estimation of the proposed hybrid quantum computing schemes with H-cat and H-coh qubits. We also compare the resource efficiencies of proposed schemes with existing MBQC photonic schemes based on direct encoding \cite{Herrera2010}, repetition code \cite{Omkar2022}, and parity encoding of multiple photons \cite{LeeSH2023} as well as circuit-based photonic schemes \cite{Dawson2006,Lund2008,Cho2007,Hayes2010,LeeSW2015}. For resource analysis of other schemes, we refer to Refs. \cite{Omkar2021,Omkar2022}. It shows that our hybrid approach is at least an order of magnitude more resource-efficient compared to all the other photonic proposals with respect to the cost of the resources, while the direct comparison is not straightforward due to the different types of resource states. The improved resource efficiency is achieved in our scheme thanks to the deterministic nature of the proposed hybrid fusion and the loss-tolerance of the resource states by inherently encoded bosonic cat-code.

\section{Physical implementations}

\subsection{\label{sec:Resource}Resource state generation}

\begin{figure}[t]
        \centering \includegraphics[clip=true, width=\columnwidth]{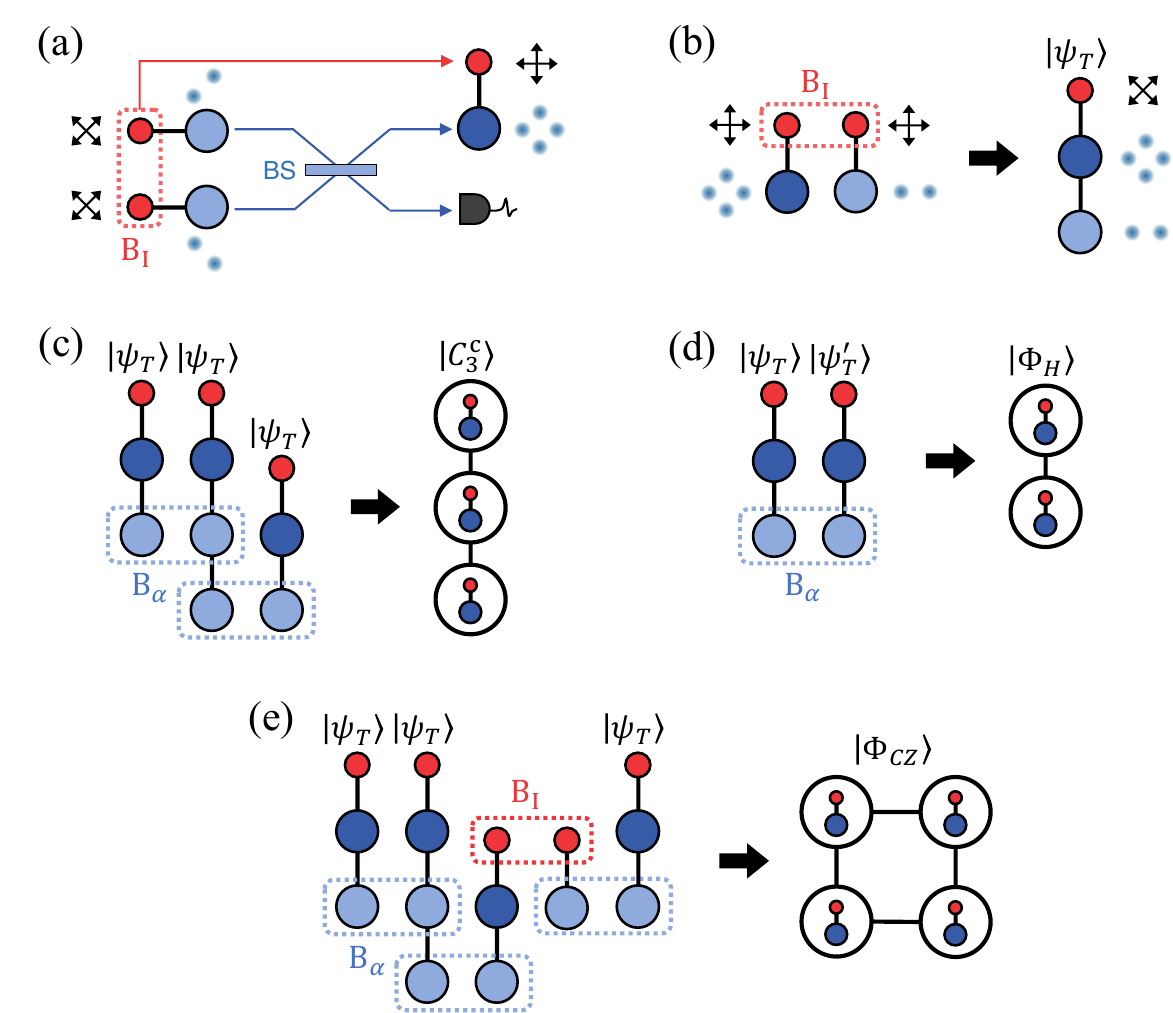}
        \caption{(a) Generation scheme of a H-cat pair. Red circles represent DV qubits and its polarization bases are represented as arrows. Blue circles represent cat-code qubits using four components of coherent states and light blue circles represent coherent-state qubits using two components of coherent states. (b) Generation scheme of a DV-coherent-cat triple $\ket{\psi_T}$. Another type of triple $\ket{\psi_T'}$ can be obtained by rotating the DV basis of H-coh pair from $H/V$ to $+/-$. By merging hybrid triples and hybrid pairs, we can generate off-line resource states such as (c) 3-qubit H-cluster state $\ket{C_3^\mathrm{c}}$, (d) Hadamard resource $\ket{\Phi_H}$, and (e) controlled-Z resource $\ket{\Phi_{CZ}}$.}
        \label{fig:Resource}
    \end{figure}

We introduce a H-coh pair $ ( \ket{+}\ket{\alpha} + \ket{-}\ket{-\alpha} )/\sqrt{2} $ as a unit resource, which is employed as a logical qubit in previous hybrid approaches \cite{LeeSW2013,Omkar2020}. It can be prepared by the scheme of Refs.~\cite{Morin2014,Guccione2020,Darras2023}, where DV-CV entanglement is heralded by the detection of single photon after a weak beamsplitter interaction between CV and DV modes. Since this scheme employs the CV basis using two coherent states $\ket{\alpha}$ and $\ket{-\alpha}$, we have to extend it to four-component cat code using $\ket{\alpha}$, $\ket{i\alpha}$, $\ket{-\alpha}$, and $\ket{-i\alpha}$. The generation of four-component cat state was proposed in Ref.~\cite{Hastrup2020}, where a pair of two-component cat state in different phases are mixed by a beamsplitter interaction followed by photon number counting on one of the output modes. Here we propose the generation of a H-cat pair where the scheme is described in Fig.~\ref{fig:Resource}(a). We begin with two H-coh pairs, $ ( \ket{+}\ket{\omega\alpha} + \ket{-}\ket{-\omega\alpha} ) \otimes ( \ket{+}\ket{-i\omega\alpha} + \ket{-}\ket{i\omega\alpha} ) $ where $ \omega=e^{i\pi/4} $. For convenience, we omit the normalization factors. DV qubits are merged by type I fusion operation $\mathrm{B}_\mathrm{I}$ and coherent-state qubits are mixed by a 50/50 beamsplitter, resulting in the state $ \ket{H}\left( \ket{\alpha}\ket{i\alpha} + \ket{-\alpha}\ket{-i\alpha} \right) +\ket{V}\left( \ket{i\alpha}\ket{\alpha} + \ket{-i\alpha}\ket{-\alpha} \right) $. We count the photon number on one of the CV modes and postselect even photon detection to rule out odd cat states. Then we have the state $ i^n\ket{H}\ket{\mathcal{C}_{\alpha}^+} + \ket{V}\ket{\mathcal{C}_{i\alpha}^+} $ and the phase $i^n$ can be compensated by $Z$ gate when $ n = 4k+2 $. Note the the DV basis can be freely interchanged between $H/V$ and $+/-$ using a waveplate. The success probability of $\mathrm{B}_\mathrm{I}$ is 50\% and the probability to detect even photons are approximately 50\% when $ \alpha\gg1 $. Therefore, we need approximately 8 H-coh pairs on average to generate a H-cat pair. Note that any imperfections and loss during the process cause to cost few more H-cat qubits on average. If we also take into account such additional errors during the generation process of H-cat qubits, the loss thresholds can be slightly degraded in Fig.~\ref{fig:RHG}(b).

To generate cluster states of hybrid qubits, we employ ancilla qubits which will be merged by fusion operation. We here choose coherent-state qubits as ancilla, and therefore we prepare DV-cat-coherent triples $ \ket{\psi_T} = \ket{+}\ket{\mathcal{C}_{\alpha}^+}\ket{\beta} + \ket{-}\ket{\mathcal{C}_{i\alpha}^+}\ket{-\beta} $ using the scheme shown in Fig.~\ref{fig:Resource}(b). The amplitude of coherent-state qubit $\beta$ does not need to be the same as the amplitude of cat-code qubit $\alpha$, but $ \beta\gtrsim1 $ is required to suppress the failure probability of fusion operation. We merge a H-cat pair $ \ket{H}\ket{\mathcal{C}_{\alpha}^+} + \ket{V}\ket{\mathcal{C}_{i\alpha}^+} $ with a H-coh pair $ \ket{H}\ket{\beta} + \ket{V}\ket{-\beta} $ using $\mathrm{B}_\mathrm{I}$ and obtain the state $\ket{\psi_T}$. In the generation of $\ket{\psi_T}$, we need $ 2\times(8+1)=18 $ H-coh pairs on average. In our scheme, we need resource states which are equivalent to cluster states with Hadamard gates applied on some qubits. For that purpose, we prepare a different type of triple state $ \ket{\psi_T'} = \ket{+}\ket{\mathcal{C}_{\alpha}^+}(\ket{\beta}+\ket{-\beta}) + \ket{-}\ket{\mathcal{C}_{i\alpha}^+}(\ket{\beta}-\ket{-\beta}) $, which can be obtained by rotating the DV basis of H-coh pair from $H/V$ to $+/-$. Coherent-state qubits are consumed by fusion operation when generating offline resource states. The fusion operation is performed by the Bell measurement $\mathrm{B}_\alpha$, which can be implemented using one beamsplitter and two PNR detectors \cite{LeeSW2013,Jeong2001}. It fails if both PNR detectors click no photon, which occurs with the probability is $ p_\alpha = \exp(-2|\beta|^2) $.

In the construction of RHG lattice, we use two kinds of 3-qubit micro H-cluster states, $\ket{C_3^\mathrm{c}}$ and $\ket{C_3^\mathrm{s}}$. Typical 3-qubit cluster states can be represented as $ \ket{C_3}_{123} = CZ_{12}CZ_{23}\ket{+_1, +_2, +_3} $. Micro cluster states used in our scheme differ by Hadamard gates from typical cluster states. A central micro cluster state is given by $ \ket{C_3^\mathrm{c}}_{102} = H_1 H_2\ket{C_3}_{102} = (\ket{0_1, 0_0, 0_2} + \ket{1_1, 1_0, 1_2})/\sqrt{2} $ and a side micro cluster state is given by $ \ket{C_3^\mathrm{s}} = H_1\ket{C_3}_{102} = (\ket{0_1, 0_0, 0_2} + \ket{1_1, 1_0, 0_2} + \ket{0_1, 0_0, 1_2} - \ket{1_1, 1_0, 1_2})/2 $. The subscript 0 represents a data qubit which will be the vertex of RHG lattice, while the subscripts 1 and 2 are for side qubits which will be consumed by hybrid fusion. The generation of $\ket{C_3^\mathrm{c}}$ is described in Fig.~\ref{fig:Resource}(c). We prepare three hybrid triples $\ket{\psi_T}$, where the second one has two coherent-state qubits which can be obtained by splitting larger coherent-state qubit with the amplitude $\sqrt{2}\beta$. Then we perform Bell measurements $\mathrm{B}_\alpha$ on coherent-state qubits to merge three hybrid qubits. We postselect on successful Bell measurements only so that no error is induced on off-line resource states. The desired state $\ket{C_3^\mathrm{c}}$ is obtained after applying the appropriate Pauli corrections according to the measurement outcome of $\mathrm{B}_\alpha$. The generation scheme of $\ket{C_3^\mathrm{s}}$ is almost the same, but one needs to replace the last triple state with $\ket{\psi_T'}$. The average number of unit resources to generate a three-qubit cluster state is $ 54(1-p_\alpha)^{-2} $.

The generation of resource states for gate teleportation, $\ket{\Phi_H}$ and $\ket{\Phi_{CZ}}$, is described in Fig.~\ref{fig:Resource}(d) and (e), respectively. The Hadamard resource $\ket{\Phi_H}$ can be generated by merging $\ket{\psi_T}$ and $\ket{\psi_T'}$ using $\mathrm{B}_\alpha$. For the controlled-Z resource $\ket{\Phi_{CZ}}$, we need to prepare another type of resource which has two coherent-state ancillae with different basis. To prepare such a state, we merge a triple $ \ket{H}\ket{\mathcal{C}_{\alpha}^+}(\ket{\beta}+\ket{-\beta}) + \ket{V}\ket{\mathcal{C}_{i\alpha}^+}(\ket{\beta}-\ket{-\beta}) $ and a H-coh pair $ \ket{H}\ket{\beta} + \ket{V}\ket{-\beta} $ using $\mathrm{B}_\mathrm{I}$. Once $\mathrm{B}_\mathrm{I}$ is successful, we merge the state with three $\ket{\psi_T}$'s using $\mathrm{B}_\alpha$. The average number of unit resources to generate $\ket{\Phi_H}$ and $\ket{\Phi_{CZ}}$ is, respectively, $ 36(1-p_\alpha)^{-1} $ and $ ( 18\times3 + (18+1)\times2 )(1-p_\alpha)^{-3} = 92(1-p_\alpha)^{-3} $.

\subsection{\label{sec:Implementation}Implementation in other platforms}

A DV-CV hybrid entangled pair is employed in our approach as the basic building block of hybrid quantum computation. In CV part, a bosonic cat-code state needs to be implemented and entangled with a DV qubit such as single photon. While we focus on all-optical implementation in our analysis, we also emphasize that our hybrid approach is applicable to other platforms. CV qubits in cat states have been successfully demonstrated in superconducting circuits \cite{Leghtas2015,Ofek2016,Lescanne2020,Grimm2020,Vlastakis2013,Wang2016,Wang2022,He2023,Pan2023}, ion traps \cite{Gan2020,Eickbusch2022,Monroe2021,Heeres2017,Blais2021,Monroe1996,Kienzler2016,Johnson2017,Jeon2024}, and circuit acoustic devices \cite{Chu2018,Hann2019,Bild2023}. Especially, the error correction using cat code has been demonstrated in a superconducting circuit system \cite{Ofek2016}.
Another line of efforts in the superconducting circuit system is a stabilized cat qubit by two-photon driven dissipation \cite{Leghtas2015,Touzard2018,Lescanne2020,Gertler2023} or Kerr nonlinear interaction \cite{Puri2017,Puri2019,Puri2020,Grimm2020,Putterman2022,Gravian2023}. By stabilizing the cat-code subspace, a bit-flip error rate is exponentially suppressed so that it suffices to correct phase-flip error using simple error correction code with low overhead \cite{Guillaud2019,Guillaud2021,Chamberland2022,Regent2023,Gouzien2023}. In those schemes, the implementation of universal gate set preserving the error bias enables hardware-efficient fault-tolerant quantum error correction \cite{Guillaud2019,Chamberland2022,Puri2020}.
In the presence of CV qubits, any interaction strong enough to entangle the CV qubits with other DV qubits can produce the hybrid qubits. For example, H-cat qubits can be generated by applying a cross-Kerr interaction between DV and CV qubits, acting as $ (\ket{+}+\ket{-})\ket{\mathcal{C}_{\alpha}^+} \to \ket{+}\ket{\mathcal{C}_{\alpha}^+} + \ket{-}\ket{\mathcal{C}_{i\alpha}^+} = \ket{+_L} $. 

While the required nonlinearity is also within the reach of the current photonic technology \cite{Sagona2020,Cui2022}, in other platforms, strong nonlinear interactions between DV and CV qubits are readily available. In trapped-ion systems, a variety of hybrid gate operations have been demonstrated, coupling the internal states of atoms and the phonon modes \cite{Monroe2021,Gan2020,Chen2023,Katz2023}. In superconducting circuit QED, universal control of the oscillator mode in a microwave cavity was demonstrated using nonlinear interaction between the oscillator mode and the transmon qubit \cite{Eickbusch2022,Heeres2017,Blais2021,Kudra2022}. Though those works have restricted the role of DV mode to the control qubit for manipulating CV mode, the coupling between CV mode and reliable DV qubit will provides us a resource for the hybrid quantum computation.

\section{\label{sec:Discussion}Discussion}

We introduced a scheme for fault-tolerant quantum computing based on hybrid qubits combining both CV and DV qubits. We developed a fault-tolerant architecture in photonic platforms by concatenating CV and DV quantum error correction codes. A hybrid qubit (H-cat) is defined in physical level by employing the bosonic cat-code in CV part and single photon in DV part. The effect of photon loss is readily correctable via the cat-code in each H-cat qubit, and the error caused by non-orthogonality is removed. We proposed two types of hybrid fusion schemes as the building block for both circuit-based and measurement-based quantum computing, to design fault-tolerant architecture based on outer DV quantum error correction code such as Steane \cite{Steane1996} and surface codes \cite{Raussendorf2006,Raussendorf2007,Fowler2009} implemented in RHG lattice. We have shown that the proposed hybrid approach allows us to achieve 4-times higher loss thresholds than existing CV and hybrid approaches and is at least an order of magnitude more resource-efficient over previous proposals in photonic platforms. 

Bosonic encoding with coherent or cat states in CV qubits generally exhibits a tradeoff between the $X$ error and $Z$ error rates, which are induced dominantly from the ambiguity of logical basis and photon loss, respectively. Note that any cat states are more fragile to photon loss, i.e.,~$Z$ error rate increases as getting larger amplitude $\alpha$, while the basis become more distinguishable to each other, i.e.,~$X$ error rate decreases \cite{Lund2008,LeeSW2013,Omkar2020,Omkar2022}. In this circumstance, the proposed hybrid scheme aims to reduce the $X$ error rate significantly by increasing the amplitude $\alpha$ while suppressing the increase of $Z$ error rate by means of the bosonic cat-code encoding on the physical level. This is how our scheme achieves the record-high loss thresholds among the schemes using cat-state encoding or their hybrid. 
Moreover, although H-cat qubit is 8 times as expensive as H-coh qubit in photonic platforms as we estimated in Sec.~\ref{sec:Resource}, we demonstrated that employing the bosonic cat-code in physical level (H-cat) can also improve the resource efficiency about 13 times compared to only coherent- or cat-state encoding (H-coh) in hybrid quantum computing. 
The resource efficiency of our scheme is due to that H-cat qubits inherit the benefit of bosonic CV quantum error correction.
In this sense a comparison of our scheme with the MBQC with cat-code only is further expected, which has not been so far rigorously studied and is currently being investigated in photonic platform \cite{Kang}.

In our analysis of the thresholds, the error model in physical level has been estimated mainly based on the effect of photon loss, which is the dominant source of errors in photonic platform. We note that other type of errors such as phase shift is hardly caused by hybrid fusion operations, i.e.,~in beamsplitters and photon-number-counting detectors.
On the other hand, phase error that may occur in logical gate operations is detrimental and should be suppressed. A set of bias-preserving gate operations can be considered to prevent further propagating of errors to other hybrid qubits \cite{Guillaud2019,Chamberland2022,Puri2020}. A controlled rotation can be also employed to suppress phase error if enough nonlinear interaction is available as illustrated in Ref.~\cite{Grimsmo2020}.
Further, in the fault-tolerance architecture, any errors in logical level caused from physical level or universal gate operations are handled by outer quantum error correction codes, e.g, the surface code in RHG lattice in our simulation.

To realize full fault-tolerant quantum computing with H-cat qubits, encoding with amplitude $\alpha \gtrsim 2$ in CV part is required, while its maximum performance is achieved with $\alpha \approx 2.93$, as presented in Fig.~\ref{fig:RHG}(b). Despite being still challenging in photonic platforms, there have been a lot progresses in generating cat states with large amplitude both experimentally \cite{Ourjoumtsev2007,Sychev2017,Hacker2019} and theoretically  \cite{Lund2004,Thekkadath2020,Takase2021,Li2023}. 
The photon number resolving (PNR) detector is another requirement in the realization of our scheme. The maximum performance of our scheme requires the resolution $ \bar{n} = \alpha^2 \approx 8.58 $, which are within the reaches of currently available detectors with the resolution up to 16 photons and 98\% efficiency \cite{Morais2020}. We thus expect that our hybrid scheme will be available with current and near-term devices.

While we have discussed mainly hybrid quantum computation in all-photonic platforms, we emphasize again that our scheme is applicable to generally other hybrid platforms including superconducting and trapped-ion systems, details of which is discussed in Sec.~\ref{sec:Implementation}. It may be further valuable to extend our scheme to combine with the fault-tolerant architecture concatenated with bosonic cat-code in superconducting systems~\cite{Chamberland2022}. Considering the biased errors under loss in cat-state encoding in CV part, other DV quantum error correction codes may be also useful for concatenation in architecture such as repetition~\cite{Guillaud2019,Chamberland2022,Gouzien2023}, and XZZX~\cite{Ataides2021,Darmawan2021,Xu2023a} codes to enhance the loss-thresholds as well as resource-efficiency further. Another interesting way to pursue is to explore hybrid encoding using squeezed-cat qubits \cite{Schlegel2022,Xu2023b,Hillmann2023,Pan2023}, which has favorable error-bias properties with reduced excitation number and partial correction of excitation loss errors.
As the thresholds have been widely investigated for stabilized cat qubits \cite{Darmawan2021,Chamberland2022,Gouzien2023,Regent2023} and for GKP qubits \cite{Fukui2017,Fukui2018,Noh2020,Noh2022} in superconducting systems, their hybrid approach incorporating DV qubits would be expected to improve the performance. For the hybridization between CV and DV qubits, nonlinear interactions can play an important role, which are readily available in superconducting or trapped-ion systems \cite{Monroe2021,Gan2020,Chen2023,Katz2023,Eickbusch2022,Heeres2017,Blais2021,Kudra2022}.
A hybrid architecture compromising GKP qubits and squeezed vacuum states can be also considered in photonic platform~\cite{Bourassa2021}.

The hybrid approach using CV-DV hybrid qubits are also generally useful and can be extended further for other quantum information processing tasks, e.g., quantum communications or sensing, which we leave as future works. We believe that our work opens a novel way to bring a synergy between CV and DV hybrid platforms towards scalable fault-tolerant quantum computing.

\section*{Acknowledgement}
This research was funded by Korea Institute of Science and Technology (2E32241) and National Research Foundation of Korea (2022M3K4A1094774). S.H.L. and H.J. are supported by National Research Foundation of Korea (NRF) grants funded by the Korean government (Grant~Nos. 2023R1A2C1006115 and 2022M3K4A1097117) and by the Institute of Information \& Communications Technology Planning \& Evaluation (IITP) grant funded by the Korea government (MSIT) (IITP-2021-0-01059 and IITP-2023-2020-0-01606). L.J. acknowledges support from the ARO(W911NF-23-1-0077), ARO MURI (W911NF-21-1-0325), AFOSR MURI (FA9550-19-1-0399, FA9550-21-1-0209, FA9550-23-1-0338), NSF (OMA-1936118, ERC-1941583, OMA-2137642, OSI-2326767, CCF-2312755), NTT Research, and the Packard Foundation (2020-71479).

\bibliography{HQC_bib}

\clearpage

\appendix

\section{\label{sec:Fusion}Hybrid fusion}

The hybrid fusion operation is a key ingredient for our hybrid quantum computation, which implements photon loss correction as well. In this section, we present the details of hybrid fusion. The fusion is composed of two Bell measurements, one for DV qubits and the other for cat-code qubits.

\subsection{\label{sec:FusionHybrid}Hybrid fusion operation}

The hybrid fusion operation can be done by performing Bell measurement respectively on CV qubits and DV qubits, that is, $\mathrm{B}_\mathcal{C}$ and $\mathrm{B}_\mathcal{D}$. Hybrid Bell states can be represented, in the basis of Bell states of CV and DV qubits, as
    \begin{eqnarray}
        \ket{\Psi_L^\pm} & = & \frac{1}{\sqrt{2}}\left( \ket{+,\mathcal{C}_{\alpha}^+}\ket{-,\mathcal{C}_{i\alpha}^+} \pm \ket{-,\mathcal{C}_{i\alpha}^+}\ket{+,\mathcal{C}_{\alpha}^+} \right) \nonumber \\
        & = & \frac{1}{2}\left[ \left( \ket{\Phi_\mathcal{D}^-} + \ket{\Psi_\mathcal{D}^-} \right) \left( \frac{\ket{\Psi_\mathcal{C}^+}}{2\mathcal{N}_{C}^{B+}} + \frac{\ket{\Psi_\mathcal{C}^-}}{2\mathcal{N}_{C}^{B-}} \right) \right. \nonumber \\
        & & \hspace{25pt} \left. \pm\left( \ket{\Phi_\mathcal{D}^-} - \ket{\Psi_\mathcal{D}^-} \right) \left( \frac{\ket{\Psi_\mathcal{C}^+}}{2\mathcal{N}_{C}^{B+}} - \frac{\ket{\Psi_\mathcal{C}^-}}{2\mathcal{N}_{C}^{B-}} \right) \right] \nonumber \\
        & = & \frac{1}{2}\left( \ket{\Phi_\mathcal{D}^-}\ket{\tilde{\Psi}_\mathcal{C}^\pm} + \ket{\Psi_\mathcal{D}^-}\ket{\tilde{\Psi}_\mathcal{C}^\mp} \right)
    \end{eqnarray}
and similarly,
    \begin{equation}        
        \ket{\Phi_L^\pm} = \frac{1}{2}\left( \ket{\Phi_\mathcal{D}^+}\ket{\tilde{\Phi}_\mathcal{C}^\pm} + \ket{\Psi_\mathcal{D}^+}\ket{\tilde{\Phi}_\mathcal{C}^\mp} \right) .
    \end{equation}
Let us first consider that the CV Bell state is determined without ambiguity. Then, the letter of hybrid Bell state is determined the same as that of CV. To determine the sign, we only need the letter information of DV Bell state, that is, the sign of CV and hybrid is the same (flipped) if the letter of DV is $\Phi$($\Psi$). Even though the DV Bell measurement $\mathrm{B}_\mathcal{D}$ fails, the letter of DV is perfectly distinguished, and therefore we can always determine the hybrid Bell state without ambiguity.

\begin{figure*}[t] 
        \centering \includegraphics[clip=true, width=0.9\linewidth]{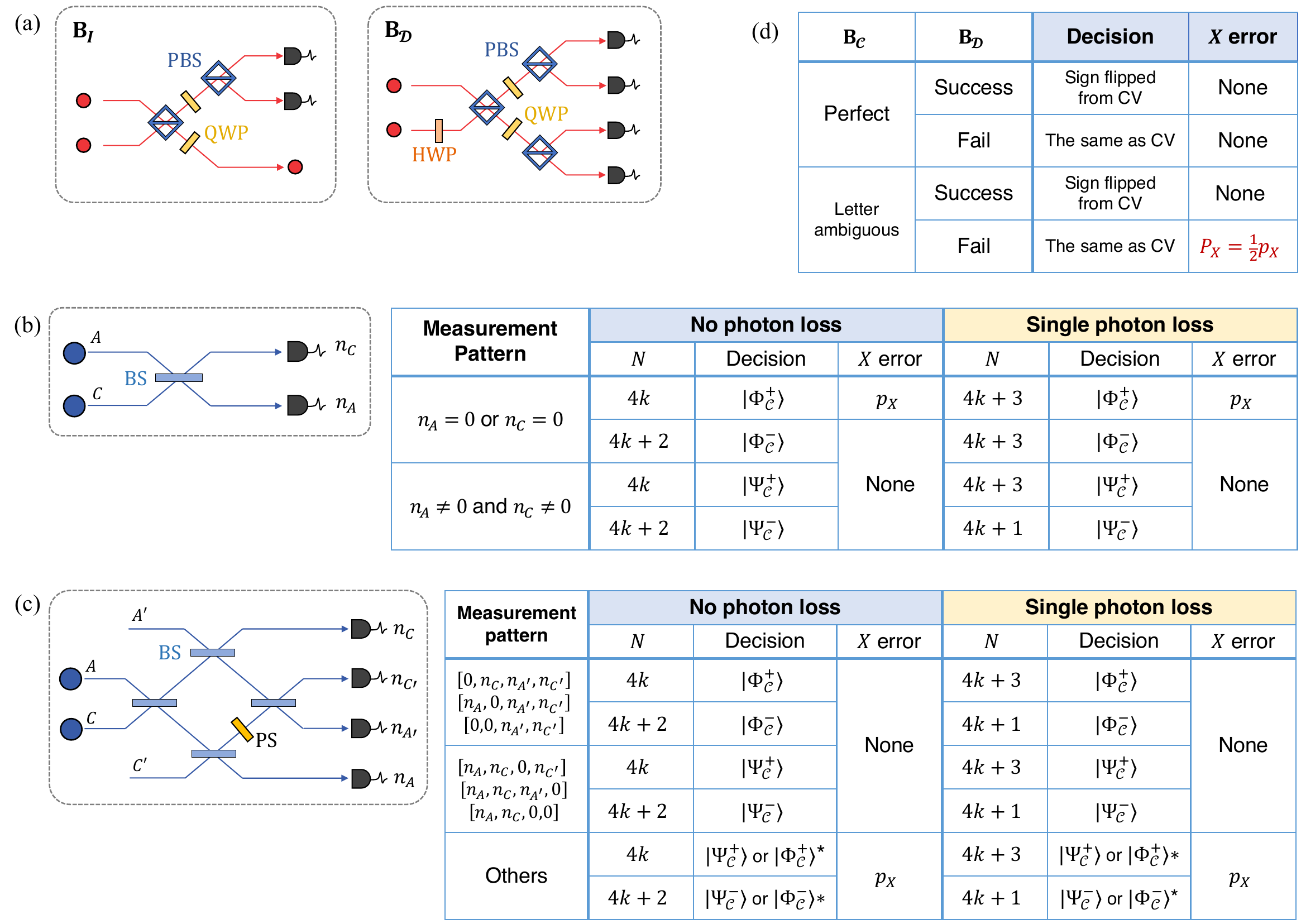}
        \caption{(a) Schemes of type I fusion operation $\mathrm{B}_\mathrm{I}$ and type II fusion operation $\mathrm{B}_\mathcal{D}$. PBS, QWP, and HWP represent polarizing beamsplitter, quater waveplate, and half waveplate, respectively. On-off detectors suffice for Bell measurements in ideal cases, while detectors are required to resolve up to two photons in order to distinguish photon loss cases. (b) Scheme for Bell measurement of cat-code qubits using one beamsplitter (BS) and two PNR detectors \cite{Hastrup2022}. The measurement outcome is determined following to the table. (c) Scheme for Bell measurement of cat-code qubits using four BSs, one $\frac{\pi}{2}$ phase shifter (PS) and four PNR detectors \cite{Su2022}. Two ancilla modes $A'$ and $C'$ are prepared in vacuum. The decision of measurement outcome is summarized in the table. We explicitly indicate $n_j$ where $ j \in \{ A, C, A', C' \} $ other than 0 if the detector clicks at least one photon. (d) The outcome decision of hybrid fusion operation. The decision is made by collecting the information of $\mathrm{B}_\mathcal{C}$ and $\mathrm{B}_\mathcal{D}$.}
        \label{fig:Fusion}
    \end{figure*}

If there is a letter ambiguity in CV Bell measurement, the ambiguity is removed if we have the sign information of DV, that is, $\mathrm{B}_\mathcal{D}$ succeds. The ambiguity only remains when $\mathrm{B}_\mathcal{C}$ has an ambiguity and $\mathrm{B}_\mathcal{D}$ fails, and it causes $X$ error with the rate $ P_X = \frac{1}{2}p_X $. We summarize the outcome decision of hybrid fusion in Fig.~\ref{fig:Fusion}(d).

\subsection{\label{sec:FusionCV}Fusion in CV part}

Cat code is designed to protect qubits against photon loss \cite{Leghtas2013,Mirrahimi2014}, which is the dominant type of error in optical systems. The key idea of four-component cat code is that we encode qubits in the subspace of even photon number states and an error is detected when the parity changes. Specifically, the code space is defined by even cat states $ \left\{ \ket{\mathcal{C}_{\alpha}^+} , \ket{\mathcal{C}_{i\alpha}^+} \right\} $, where $ \ket{\mathcal{C}_{\alpha}^+} = \mathcal{N}_\mathcal{C}^+ (\ket{\alpha}+\ket{-\alpha}) $ with the normalization factor $ \mathcal{N}_\mathcal{C}^+ \equiv 1/\sqrt{2(1+e^{-2\alpha^2})} $ and $ \ket{\alpha} $ is a coherent state with the amplitude $ \alpha $. A photon loss can be modeled by application of the annihilation operator so that the state evolves into the odd cat states as $ \hat{a} \ket{\mathcal{C}_{\alpha}^+} \propto \ket{\mathcal{C}_{\alpha}^-} = \mathcal{N}_\mathcal{C}^- (\ket{\alpha}-\ket{-\alpha}) $ where $ \mathcal{N}_\mathcal{C}^- \equiv 1/\sqrt{2(1-e^{-2\alpha^2})} $ is the normalization factor. Further, photon loss induces a phase shift on $ \ket{\mathcal{C}_{i\alpha}^+}$, i.e., $ \hat{a} \ket{\mathcal{C}_{i\alpha}^+} \propto i\ket{\mathcal{C}_{i\alpha}^-} $, which may cause $Z$ error unless the parity change is detected by the measurement. Thus a parity measurement can play the role of syndrome measurement to detect a photon loss event. Such measurement has been demonstrated in the superconducting circuit system via nonlinear interaction with an ancilla transmon qubit \cite{Ofek2016}.
    
In optical systems, two separate works have recently proposed schemes for Bell measurement of four-component cat code \cite{Su2022,Hastrup2022}, which detect single photon loss as well. The cat-code Bell measurement distinguishes four Bell states $ \ket{\Psi_\mathcal{C}^\pm} = \mathcal{N}_\mathcal{C}^{B\pm}( \ket{\mathcal{C}_{\alpha}^+}\ket{\mathcal{C}_{i\alpha}^+} \pm \ket{\mathcal{C}_{i\alpha}^+}\ket{\mathcal{C}_{\alpha}^+}) $, $ \ket{\Phi_\mathcal{C}^\pm} = \mathcal{N}_\mathcal{C}^{B\pm}( \ket{\mathcal{C}_{\alpha}^+}\ket{\mathcal{C}_{\alpha}^+} \pm \ket{\mathcal{C}_{i\alpha}^+}\ket{\mathcal{C}_{i\alpha}^+}) $ with the normalization factor $ \mathcal{N}_\mathcal{C}^{B\pm} = \cosh\alpha^2 / \sqrt{2(\cosh^2\alpha^2\pm\cos^2\alpha^2)} $. The schemes only use linear optical elements and PNR detectors, as depicted in Fig.~\ref{fig:Fusion}(b,c). Measurement patterns on PNR detectors differ by the parity of input ports. In what follows, we will use the notation \{even(odd),even(odd)\} to represent the photon number parity of input mode $A$ and $C$ and define $N$ to be the total number of photons detected.

\subsubsection{Scheme by Hastrup \textit{et al.} \cite{Hastrup2022}}

The scheme proposed by Hastrup and Anderson (HA) is composed of one beamsplitter and two PNR detectors, as depicted in Fig.~\ref{fig:Fusion}(b).The transformation of coherent states by beamsplitter is described as
    \begin{equation} \label{eq:UBellCatHA}
        \ket{\alpha_A}_A\ket{\alpha_C}_C\rightarrow \Big|\frac{\alpha_A+\alpha_C}{\sqrt{2}}\Big\rangle_A \Big|\frac{\alpha_A-\alpha_C}{\sqrt{2}}\Big\rangle_C .
    \end{equation}
PNR detectors count the photon numbers of modes $A$ and $C$, which we denote by $n_A$ and $n_C$, respectively. If the input state is a superposition of $ \ket{\mathcal{C}_{\alpha}^+}\ket{\mathcal{C}_{\alpha}^+} $ and $ \ket{\mathcal{C}_{i\alpha}^+}\ket{\mathcal{C}_{i\alpha}^+} $, where every term in the coherent-state basis is given by either $ \alpha_A=\alpha_C $ or $ \alpha_A=-\alpha_C $, at least one of detectors must click no photon. In other words, if $ n_A=0 $ or $ n_C=0 $, we guess that the letter of input Bell state is $\Phi$, and otherwise, we guess that the letter is $\Psi$. On the other hand, the sign can be distinguished by the total number of photon detected $N$. In the ideal case without photon loss where the input parity is \{even,even\}, a simple calculation yields that the sign is determined as $+$ for $ N=4k $ and $-$ for $ N=4k+2 $ where $k$ is a nonnegative integer.

Let us consider the case that single photon is lost from one of the modes $A$ or $C$, where the input parity becomes \{odd,even\} or \{even,odd\}. The photon loss only changes the relative phase between superposed coherent states, but does not change the amplitude of coherent states in phase space. Therefore, we can apply the same strategy of letter decision. Due to single photon loss, the sign is determined as $+$ for $ N=4k+3 $ and $-$ for $ N=4k+1 $. The outcome decision is summarized in the table of Fig.~\ref{fig:Fusion}(b), for photon loss case as well as for ideal case.

If single photon is lost from each input port so that the input parity is \{odd,odd\}, we follow the outcome decision of ideal case because $N$ is even. $Z$ error is induced by a phase shift on $ \ket{\mathcal{C}_{i\alpha}^+} $ and it can be corrected only if the measurement pattern distinguishes whether the input parity is \{even,even\} or \{odd,odd\}. In the case that both $n_A$ and $n_C$ are nonzero, a straightforward calculation yields that $ n_C = n_A ~(\textrm{mod } 4) $ for \{even,even\} input and $ n_C = n_A+2 ~(\textrm{mod } 4) $ for \{odd,odd\} input, and thus $Z$ error is correctable. Otherwise, $Z$ error remains undetected, of which the probability is given by
    \begin{equation}
        p_{Z|OO}(\alpha) = \frac{1}{2} + \frac{1}{2}\csch^2\alpha^2( \cosh\alpha^2 - \cos\alpha^2 ) ,
    \end{equation}
where $OO$ in the subscript represents that $Z$ error occurs for \{odd,odd\} input.

There exists small probability of misidentifying $\Psi$ as $\Phi$ because no photon click can occur for a coherent state with nonzero amplitude. Once we obtain the measurement pattern with $ n_A=0 $ or $ n_C=0 $, we have an ambiguity in the letter, which causes $X$ error. We can locate such $X$ error and denote by $p_\textrm{loc}$ and $p_X$, respectively, the probability to locate $X$ error and the total $X$ error rate. In general, the probability depends on the input state, but in our hybrid quantum computation, it suffices to calculate the probability for the basis states $ \ket{\mathcal{C}_{\alpha}^\pm} $ and $ \ket{\mathcal{C}_{i\alpha}^\pm} $ because they are accompanied by orthogonal DV qubit states $ \ket{+} $ and $ \ket{-} $, respectively. Even though DV qubit is lost, CV qubit remains in a mixture of $ \ket{\mathcal{C}_{\alpha}^\pm} $ and $ \ket{\mathcal{C}_{i\alpha}^\pm} $ due to dephasing. We assume that the input state is equally distributed among $ \ket{\mathcal{C}_{\alpha}^\pm}\ket{\mathcal{C}_{\alpha}^\pm} $, $ \ket{\mathcal{C}_{\alpha}^\pm}\ket{\mathcal{C}_{i\alpha}^\pm} $, $ \ket{\mathcal{C}_{i\alpha}^\pm}\ket{\mathcal{C}_{\alpha}^\pm} $, and $ \ket{\mathcal{C}_{i\alpha}^\pm}\ket{\mathcal{C}_{i\alpha}^\pm} $. A direct calculation yields that the locatable error probability is given by
        \begin{eqnarray}
            p_{\textrm{loc}|EE}(\alpha) & = & \frac{1}{2} + \frac{1}{2}\sech^2\alpha^2( \cosh\alpha^2 + \cos\alpha^2 - 1 ) , \nonumber \\
            p_{\textrm{loc}|EO}(\alpha) & = & \frac{1}{2} + \frac{1}{2}\sech\alpha^2 , \\
            p_{\textrm{loc}|OO}(\alpha) & = & \frac{1}{2} + \frac{1}{2}\csch^2\alpha^2( \cosh\alpha^2 - \cos\alpha^2 ) . \nonumber 
        \end{eqnarray}
where $EE$, $EO$, and $OO$ in the subscript denote, respectively, the input parity \{even,even\}, \{even,odd\}, and \{odd,odd\}. Further the total $X$ error rate is written as
        \begin{eqnarray}
            p_{X|EE}(\alpha) & = & \frac{1}{2}\sech^2\alpha^2( \cosh\alpha^2 + \cos\alpha^2 - 1 ) , \nonumber  \\
            p_{X|EO}(\alpha) & = & \frac{1}{2}\sech\alpha^2 , \\
            p_{X|OO}(\alpha) & = & \frac{1}{2}\csch^2\alpha^2( \cosh\alpha^2 - \cos\alpha^2 ) . \nonumber 
        \end{eqnarray}
It is straightforward to observe that $ p_\textrm{loc} = \frac{1}{2}+p_X $ for any input parity, where the term $\frac{1}{2}$ represents that input states $ \ket{\mathcal{C}_{\alpha}^\pm}\ket{\mathcal{C}_{\alpha}^\pm} $ and $ \ket{\mathcal{C}_{i\alpha}^\pm}\ket{\mathcal{C}_{i\alpha}^\pm} $ always fall into unambiguous measurement patterns and the latter term represents the probability of misidentifying $\Psi$ as $\Phi$.

\subsubsection{Scheme by Su \textit{et al.} \cite{Su2022}}

The scheme proposed by Su, Dhand, and Ralph (SDR) is composed of two ancilla modes prepared in vacuum, four beamsplitters, one phase shifter, and four PNR detectors, as depicted in Fig.~\ref{fig:Fusion}(c).
The transformation of coherent states by linear optical elements is described as
\begin{widetext}
    \begin{equation} \label{eq:UBellCatSDR}
        \ket{\alpha_A}_A\ket{\alpha_C}_C\ket{vac}_{A'}\ket{vac}_{C'} \rightarrow \Big|\frac{\alpha_A+\alpha_C}{2}\Big\rangle_A \Big|\frac{\alpha_A-\alpha_C}{2}\Big\rangle_C \Big|\frac{-1+i}{2\sqrt{2}}\alpha_A+\frac{1+i}{2\sqrt{2}}\alpha_C\Big\rangle_{A'} \Big|\frac{1+i}{2\sqrt{2}}\alpha_A+\frac{-1+i}{2\sqrt{2}}\alpha_C\Big\rangle_{C'} ,
    \end{equation}
\end{widetext}
where $\ket{vac}$ represents a vacuum state. PNR detectors count the photon numbers of modes $A$, $C$, $A'$, and $C'$, which we denote by $n_A$, $n_C$, $n_{A'}$, and $n_{C'}$, respectively. In the basis of four-headed cat code, at least one of the detector must click no photon such that
    \begin{equation}
        \begin{cases}
            n_A=0 & \textrm{if}~\alpha_C=-\alpha_A , \\
            n_C=0 & \textrm{if}~\alpha_C=\alpha_A , \\
            n_{A'}=0 & \textrm{if}~\alpha_C=-i\alpha_A , \\
            n_{C'}=0 & \textrm{if}~\alpha_C=i\alpha_A .
        \end{cases}
    \end{equation}
Thus, if we obtain a measurement pattern with $ n_{A'}=0 $ or $ n_{C'}=0 $, we guess that the input state is a superposition of $ \ket{\mathcal{C}_{\alpha}^+}\ket{\mathcal{C}_{i\alpha}^+} $ and $ \ket{\mathcal{C}_{\alpha}^+}\ket{\mathcal{C}_{i\alpha}^+} $, that is, the letter of Bell state is $\Psi$. Similarly, if $ n_A=0 $ or $ n_C=0 $, we guess the letter of input Bell state is $\Phi$. The sign is determined as $+$ for $ N=4k $ and $-$ for $ N=4k+2 $ in the ideal case. In the case of single photon loss, the letter is determined in the same way and the sign is determined as $+$ for $ N=4k+3 $ and $-$ for $ N=4k+1 $. The decision of measurement outcome is summarized in the table of Fig.~\ref{fig:Fusion}(c).

$Z$ error induced from \{odd,odd\} input is correctable by a few measurement patterns. For the measurement pattern with $ n_A>0, n_C>0 $ and $ n_{A'} = n_{C'} = 0 $, $n_A$ and $n_C$ are related by $ n_A = n_C ~(\textrm{mod}~4) $ for \{even,even\} input, while $ n_A = n_C+2 ~(\textrm{mod}~4) $ for \{odd,odd\} input. For the measurement pattern with $ n_A = n_C = 0 $ and $ n_{A'}>0, n_{C'}>0 $, we can distinguish the input parity in a similar way.

We can distinguish one of four Bell states with certainty if no photon click occurs only on one side of detecters $ \{A,C\} $ or $ \{A',C'\} $. However, in some cases, no photon click may occur on both sides of detectors because there is chance of no photon click on other modes in coherent states with nonzero amplitude. In such cases, we have an ambiguity on the letter of Bell state. For example, for the measurement pattern with $ n_A = n_{A'} = 0 $ and $ n_C>0, n_{C'}>0 $, the projected state is $ 2^{-n_C/2}\ket{\Psi_\mathcal{C}^\pm} + 2^{-n_{C'}/2}\ket{\Phi_\mathcal{C}^\pm} $. If $ n_{C'}>n_C $, we take the state with higher probability $ \ket{\Psi_\mathcal{C}^\pm} $ assigning $X$ error rate $ p_X = \frac{2^{-n_{C'}}}{2^{-n_C}+2^{-n_{C'}}} $, and vice versa. If $ n_C=n_{C'} $, we randomly choose one of two states $ \ket{\Psi_\mathcal{C}^\pm} $ or $ \ket{\Phi_\mathcal{C}^\pm} $ assigning $X$ error rate $ p_X = \frac{1}{2} $. The same strategy is applied for other measurement patterns as well.

The probability of uncorrectable $Z$ error is given by
\begin{widetext}
    \begin{equation}
        p_{Z|OO}(\alpha) = \csch^2\alpha^2 \left( \tfrac{1}{2}\cosh(2\alpha^2) - \cosh\alpha^2 + 2\cosh\tfrac{\alpha^2}{2} - 2\cos\tfrac{\alpha^2}{2} + \tfrac{1}{2} \right) .
    \end{equation}
The probability to obtain measurement patterns with $X$ error is written as
    \begin{eqnarray}
        p_{\textrm{loc}|EE}(\alpha) & = & \sech^2\alpha^2 \left( \cosh\tfrac{3\alpha^2}{2} - \tfrac{1}{2}\cosh\alpha^2 + \cos\tfrac{\alpha^2}{2} - \tfrac{1}{2} \right) , \nonumber \\
        p_{\textrm{loc}|EO}(\alpha) & = & \sech\frac{\alpha^2}{2} + \frac{1}{2}\sech\alpha^2 \left( \sech^2\frac{\alpha^2}{2} - 1 \right) , \\
        p_{\textrm{loc}|OO}(\alpha) & = & \csch^2\alpha^2 \left( \cosh\tfrac{3\alpha^2}{2} - \tfrac{1}{2}\cosh\alpha^2 - \cos\tfrac{\alpha^2}{2} + \tfrac{1}{2} \right) . \nonumber
    \end{eqnarray}
\end{widetext}
Since the explicit calculation of total $X$ error rate $p_X$ is not straightforward, we rather perform a numerical calculation.

\begin{figure}[t] 
        \centering \includegraphics[clip=true, width=0.8\columnwidth]{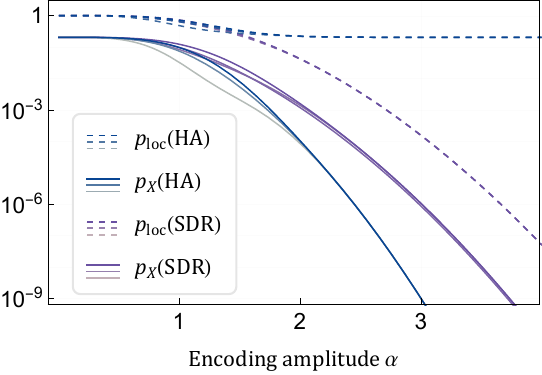}
        \caption{Plot illustrating error rates in the Bell measurement of cat-code qubits using HA scheme (blue) and SDR scheme (purple). Solid curves represent the total $X$ error rate and dashed curves represent the probability to locate $X$ error. Each curve is an overlap of three curves representing different input parity, while they are almost the same for large $\alpha$.}
        \label{fig:ErrBellCat}
    \end{figure}

We plot, in Fig.~\ref{fig:ErrBellCat}, the locatable error rate $p_\textrm{loc}$ and the $X$ error rate $p_X$ for both HA scheme and SDR scheme. The error rates are slightly different for different input parity, but they are almost overlapped especially when $ \alpha>2 $. It is shown that the total $X$ error rate exponentially decreases for both schemes, while HA scheme exhibits much lower error rate. On the other hand, the probability to locate the error is much lower for SDR scheme while it is almost constant for HA scheme even for large $\alpha$. That is because the measurement patterns to detect $\Phi$ always have an ambiguity of misidentifying $\Psi$ even though the actual error rate is very small. Therefore, the SDR scheme is beneficial in the case that we need an unambiguous discrimination of Bell state without error, while the HA scheme works better when we need to suppress the actual error rate without postselection.

\subsection{\label{sec:FusionDV}Fusion in DV part}
    
For the polarization qubit, we employ two types of Bell measurements, called type I fusion operation and type II fusion operation \cite{Browne2005}, depicted in Fig.~\ref{fig:Fusion}(a). The type I fusion operation $\mathrm{B}_\mathrm{I}$ performs a partial Bell measurement which outputs one qubit from two-qubit input. If only one is detected, the operation is described as $ \ket{+}\bra{H}\bra{H} - \ket{-}\bra{V}\bra{V} $ for $H$ click or as $ \ket{H}\bra{H}\bra{H} + \ket{V}\bra{V}\bra{V} $ for $V$ click. If two photons or no photon is detected, the measurement fail, which occurs with probability $1/2$. This operation plays a role of connecting qubits in two separate cluster states. In the resource generation described in Sec.~\ref{sec:Resource}, $\mathrm{B}_\mathrm{I}$ is employed to merge hybrid entangled states.

The type II fusion operation $\mathrm{B}_\mathrm{II}$, or $\mathrm{B}_\mathcal{D}$ in this paper, performs an incomplete Bell measurement which distinguishes only two of four Bell states. If one detector from the upper two and another from the lower two click at the same time, the measurement succeeds yielding the results $ \ket{\Psi_\mathcal{D}^+} = \frac{1}{\sqrt{2}}( \ket{H}\ket{V} + \ket{V}\ket{H} ) $ for $(H,V)$ or $(V,H)$ clicks and $ \ket{\Psi_\mathcal{D}^-} = \frac{1}{\sqrt{2}}( \ket{H}\ket{V} - \ket{V}\ket{H} ) $ for $(H,H)$ or $(V,V)$ clicks. If two photons are detected simultaneously at upper or lower detectors, the measurement fails, but we can obtain the information that the state is one of two Bell states $ \ket{\Phi_\mathcal{D}^\pm} = \frac{1}{\sqrt{2}}( \ket{H}\ket{H} \pm \ket{V}\ket{V} ) $, that is, the sign is $\Phi$. In ideal case, the failure occurs with probability $1/2$. If any of DV photon is lost, $\mathrm{B}_\mathcal{D}$ can only detect one or no photon and the information on input DV Bell state is completely lost. To distinguish between the photon loss and the failure, detectors are required to resolve up to two photons.

\section{\label{sec:GateOp}Gate operations of hybrid qubits}

\begin{figure}[t] 
        \centering \includegraphics[clip=true, width=\columnwidth]{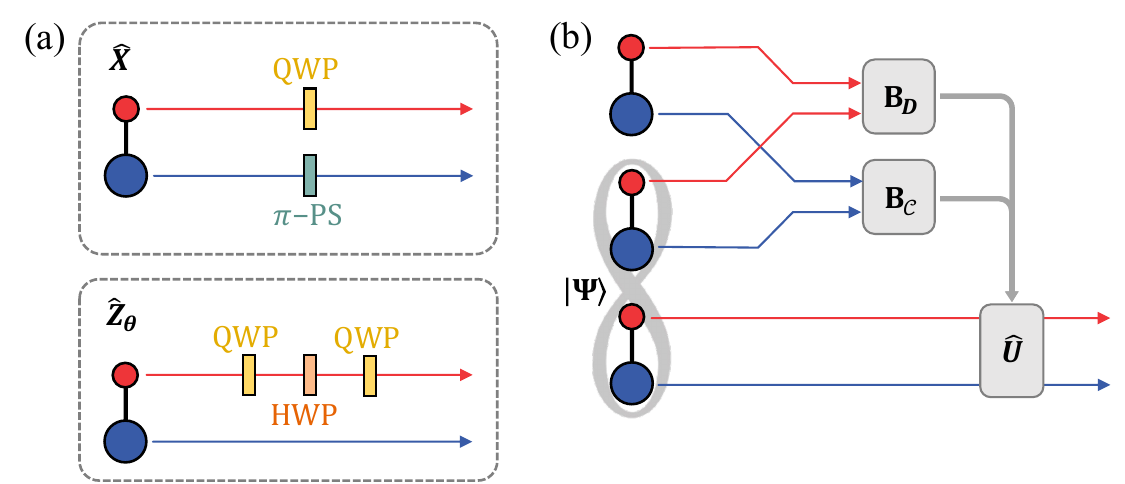}
        \caption{(a) Optical implementation of single qubit rotation gates. $\hat{X}$ can be implemented using half waveplate (HWP) on DV qubit and $\pi$-phase shifter (PS) on CV qubit. $\hat{Z}_\theta$ can be implemented using HWP sandwiched by two quater waveplates (QWPs), where the rotation angle $\theta$ is adjusted by rotating HWP. (b) Schematics for the teleportation of hybrid qubits. The measurement outcome of hybrid fusion is sent via classical channel and is used for Pauli correction.}
        \label{fig:Operation}
    \end{figure}

In this section, we present gate operations of hybrid qubits, required for universal quantum computation. In order to perform universal quantum computation, we choose the gate set $ \{ \hat{X}, \hat{Z}_\theta, \hat{H}, \hat{CZ} \} $. $\hat{X}$ and $\hat{Z}_\theta$ can be implemented only using linear optical elements, as depicted in Fig.~\ref{fig:Operation}(a). The Pauli $ X $  can be implemented by applying bit flip operations on both CV and DV qubits, using a polarization rotator acting as $ \ket{+} \leftrightarrow \ket{-} $ and a $ \frac{\pi}{2} $ phase shifter acting as $ \ket{\mathcal{C}_{\alpha}^+} \leftrightarrow \ket{\mathcal{C}_{i\alpha}^+} $. The arbitrary rotation along $ Z $ axis, $ \hat{Z}_\theta $, can be done by applying $\theta$ phase shift only on the DV qubit as $ \ket{+} \to \ket{+} , \ket{-} \to e^{i\theta}\ket{-} $, implemented by half waveplate sandwiched by two quater waveplates. To perform Hadamard and Controlled-$Z$ gates, we employ the gate teleportation technique.

The essential ingredients for the teleportation are an entangled state and the Bell measurement, as depicted in \ref{fig:Operation}(b). The hybrid Bell measurement is done by fusion operation that we have described in the previous section. We accomplish the teleportation by applying appropriate Pauli correction according to the measurement outcome of hybrid fusion. Indeed, we do not need to physically perform Pauli operations, they can be done by updating Pauli frame \cite{Dawson2006}. The gate teleportation can be down employing appropriate entangled channel. For instance, the Hadamard gate $ H $ can be performed using the resource state $ \ket{\Phi_H} \propto \ket{0_L,0_L} + \ket{0_L,1_L} + \ket{1_L,0_L} - \ket{1_L,1_L} $ and the $ CZ $ gate can be performed with two teleportation circuits using $ \ket{\Phi_{CZ}} \propto \ket{0_L,0_L,0_L,0_L} + \ket{0_L,0_L,1_L,1_L} + \ket{1_L,1_L,0_L,0_L} - \ket{1_L,1_L,1_L,1_L} $. The generation scheme of resource states is presented in the next section.

The measurement on the $Z$ basis can be performed by the measurement on DV qubit using polarization measurement. The phase measurement on CV qubit may also accomplish the $Z$-basis measurement, but requires nonlinear operations \cite{Grimsmo2020}. The heterodyne measurement can be a candidate for the cat-code measurement, taking practical advantages with the cost of a high probability of misreading.

\section{\label{sec:ErrorModel}Error Model}

\subsection{\label{sec:LossChannel}State evolution under lossy channel}

Photon losses are major obstacles in optical quantum computing. The interaction with lossy environment is described by the master equation \cite{Phoenix1990}
    \begin{equation}
        \frac{d\rho}{dt} = \gamma\sum_j\left( \hat{a}_j\rho\hat{a}_j^\dagger - \frac{1}{2}\hat{a}_j^\dagger\hat{a}_j\rho - \frac{1}{2}\rho\hat{a}_j^\dagger\hat{a}_j \right) , 
    \end{equation}
where $j$ runs over DV and CV modes and the loss rate is given by $ \eta = 1 - \exp(-\gamma t) $. We introduce the formula for the evolution in the coherent-state basis, written as \cite{Phoenix1990}
    \begin{equation}
        \ket{\alpha}\bra{\beta} \to \braket{\beta}{\alpha}^\eta\ket{\sqrt{1-\eta}\alpha}\bra{\sqrt{1-\eta}\beta} .
    \end{equation}
With this formula, a straightforward calculation leads to the expression for the evolution of an arbitrary hybrid qubit $ \ket{\psi} = a\ket{+}\ket{\mathcal{C}_{\alpha}^+} + b\ket{+}\ket{\mathcal{C}_{i\alpha}^+} $, written as
    \begin{equation}
        \rho = (1-\eta)\sum_{l=0}^3 A_l\ket{\psi'_{(l)}}\bra{\psi'_{(l)}} + \eta\left( B_0\tau_{(0)} + B_1\tau_{(1)} \right) ,
    \end{equation}
where
    \begin{eqnarray}
        & & \ket{\psi'_{(0)}} = a\ket{+}\ket{\mathcal{C}_{\alpha'}^+} + b\ket{-}\ket{\mathcal{C}_{i\alpha'}^+} , \nonumber \\
        & & \ket{\psi'_{(1)}} = a\ket{+}\ket{\mathcal{C}_{\alpha'}^-} + ib\ket{-}\ket{\mathcal{C}_{i\alpha'}^-} , \nonumber \\
        & & \ket{\psi'_{(2)}} = a\ket{+}\ket{\mathcal{C}_{\alpha'}^+} - b\ket{-}\ket{\mathcal{C}_{i\alpha'}^+} , \nonumber \\
        & & \ket{\psi'_{(3)}} = a\ket{+}\ket{\mathcal{C}_{\alpha'}^-} - ib\ket{-}\ket{\mathcal{C}_{i\alpha'}^-} , \\
        & & \tau_{(0)} = \ket{vac}\bra{vac} \otimes \left( |a|^2\ket{\mathcal{C}_{\alpha'}^+}\bra{\mathcal{C}_{\alpha'}^+} + |b|^2\ket{\mathcal{C}_{i\alpha'}^+}\bra{\mathcal{C}_{i\alpha'}^+} \right) , \nonumber \\
        & & \tau_{(1)} = \ket{vac}\bra{vac} \otimes \left( |a|^2\ket{\mathcal{C}_{\alpha'}^-}\bra{\mathcal{C}_{\alpha'}^-} + |b|^2\ket{\mathcal{C}_{i\alpha'}^-}\bra{\mathcal{C}_{i\alpha'}^-} \right) . \nonumber
    \end{eqnarray}
Note that the amplitude of coherent states is reduced to $ \alpha' = \sqrt{1-\eta}\alpha $. The coefficients are explicitly given by
    \begin{eqnarray}
        A_0 & = & \frac{\cosh\alpha'^2}{2\cosh\alpha^2}\left\{ \cosh(\eta\alpha^2) + \cos(\eta\alpha^2) \right\} , \nonumber \\
        A_1 & = & \frac{\sinh\alpha'^2}{2\cosh\alpha^2}\left\{ \sinh(\eta\alpha^2) + \sin(\eta\alpha^2) \right\} , \nonumber \\
        A_2 & = & \frac{\cosh\alpha'^2}{2\cosh\alpha^2}\left\{ \cosh(\eta\alpha^2) - \cos(\eta\alpha^2) \right\} , \nonumber \\
        A_3 & = & \frac{\sinh\alpha'^2}{2\cosh\alpha^2}\left\{ \sinh(\eta\alpha^2) - \sin(\eta\alpha^2) \right\} , \\
            B_0 & = & \frac{\cosh\alpha'^2 \cosh(\eta\alpha^2)}{\cosh\alpha^2} , \nonumber \\
            B_1 & = & \frac{\sinh\alpha'^2 \sinh(\eta\alpha^2)}{\cosh\alpha^2} .
    \end{eqnarray}
The coefficients $ A_0+A_2 = B_0 $ and $ A_1+A_3 = B_1 $ represent the probability that the CV qubit resides in the even and odd space, respectively.

When $l$ photons are lost from CV qubit, the state evolves into $ \ket{\psi'_{(l)}} \propto (I\otimes\hat{a}^l) \ket{\psi'_{(0)}} $. Because the state after 4 photon subtraction returns to the original state, that is, $ \ket{\psi'_{(4)}}$ and $\ket{\psi'_{(0)}}$ are equivalent up to normalization, the state evolves cyclically into four states with $ l=0,1,2,3 $. If DV photon is lost, we lose the phase information, which results in the mixed state $\tau_{(m)}$ where $ m=0(1) $ if even(odd) photons are lost from CV qubit. 

\subsection{\label{sec:ErrorAnalysis}Error analysis}

In Sec.~\ref{sec:FusionCV}, we discuss the the $X$ error rate $p_X$ which occurs in $ \mathrm{B}_\mathcal{C} $ for different parity of input states. Using the weight of even (odd) parity state after loss channel, $B_0$ ($B_1$), we obtain the failure probability of $ \mathrm{B}_\mathcal{C} $ for noisy input states, written as
    \begin{equation}
        p_X'(\alpha') = B_0^2 p_{X|EE}(\alpha') + B_1^2 p_{X|OO}(\alpha') + 2B_0B_1 p_{X|EO}(\alpha') .
    \end{equation}
The $X$ error rate of hybrid fusion becomes a half of $p_X'(\alpha')$ if both DV photons are detected during $\mathrm{B}_\mathcal{D}$. That is because we can remove the ambiguity in Bell measurement once $\textrm{B}_\textrm{I}$ succeeds. If any of input DV photon is lost, the $X$ error rate is just given by $p_X'(\alpha')$ because we cannot obtain any information from DV. Therefore, the total $X$ error rate of hybrid fusion after loss is written as 
    \begin{eqnarray}
        P_X & = & (1-\eta)^2 \frac{p_X'(\alpha')}{2} + (2\eta-\eta^2) p_X'(\alpha') \nonumber \\
        & = & \frac{1+2\eta-\eta^2}{2} p_X'(\alpha')
    \end{eqnarray}

Phase errors occur due to the photon loss. Let us first assume that $l_A$ photons are lost at CV mode $A$ and no photon is lost at CV mode $C$. As discussed in Sec.~\ref{sec:FusionCV}, no error occurs when $ l_A=0,1 ~(\textrm{mod}~4) $ and unlocatable $Z$ error occurs when $ l_A=2,3 ~(\textrm{mod}~4) $, where the modulo 4 is taken due to the cyclic behavior of cat code. The same analysis can be made if $l_C$ photons are lost at CV mode $C$ and no photon is lost at CV mode $A$. If photons are lost at both CV modes, by taking the phase shift induced on both modes into consideration, unlocatable $Z$ error occurs when $ l_A+l_C = 2,3 ~(\textrm{mod}~4) $. Once we recognize that the input parity is changed to \{odd,odd\} from the measurement pattern of $\mathrm{B}_\mathcal{C}$, we apply $Z$ gate to correct the phase error. It properly correct the error when $ l_A+l_C=2 ~(\textrm{mod}~4) $, but it induces an unwanted error when $ l_A+l_C=0 ~(\textrm{mod}~4) $. Taking all together, the unlocatable $Z$ error rate is written as
    \begin{eqnarray} \label{eq:PZunloc}
        P_{Z\textrm{(unloc)}} = (1-\eta)^2 & & \big[ 2(A_0+A_1)(A_2+A_3)  \\
        & & ~ + (A_1-A_3)^2 p_{Z|OO}(\alpha') \big] , \nonumber
    \end{eqnarray}
where the first term $(1-\eta)^2$ represents that neither of DV photons are lost. On the other hand, if we detect that any of DV photon is lost, a dephasing error occurs, which is modeled by locatable $Z$ error with the probability $1/2$. The rate of locatable $Z$ error is written as
    \begin{equation}
        P_{Z\textrm{(loc)}} = \frac{1}{2}\big[ 1 - (1-\eta)^2 \big] .
    \end{equation}

\subsection{\label{sec:ErrorFusion}Error propagation by fusion operation}

To understand error propagation in the cluster state, we briefly introduce the stabilizer formalism. In typical cluster states, stabilizers are given by the tensor product of Pauli $X$ of a qubit and Pauli $Z$'s of its adjacent qubits. Micro cluster states used in our scheme differ by Hadamard gates, which exchange the role of $X$ and $Z$ operators, from typical cluster states. For example, in the central micro cluster state $ \ket{C_3^\mathrm{c}}_{102} = H_1 H_2\ket{C_3}_{102} $, where Hadamard gates are applied on the side qubits 1 and 2, the stabilizers are given by $ Z_1Z_0 $, $ Z_0Z_2$, and $ X_1X_0X_2 $. If an error occurs on a qubit, it is propagated into adjacent qubits in a way that the stabilizer statistics are preserved \cite{LeeSH2023}. By following the stabilizer formalism, we find a simple propagation rule for micro cluster states: When a qubit is measured by fusion operation, the error in fusion, $X$ or $Z$, induces the same type of error on its adjacent qubit if both qubits are in the different Hadamard configuration, and otherwise, the type of error is changed.

\begin{figure}[t]
        \centering \includegraphics[clip=true, width=\columnwidth]{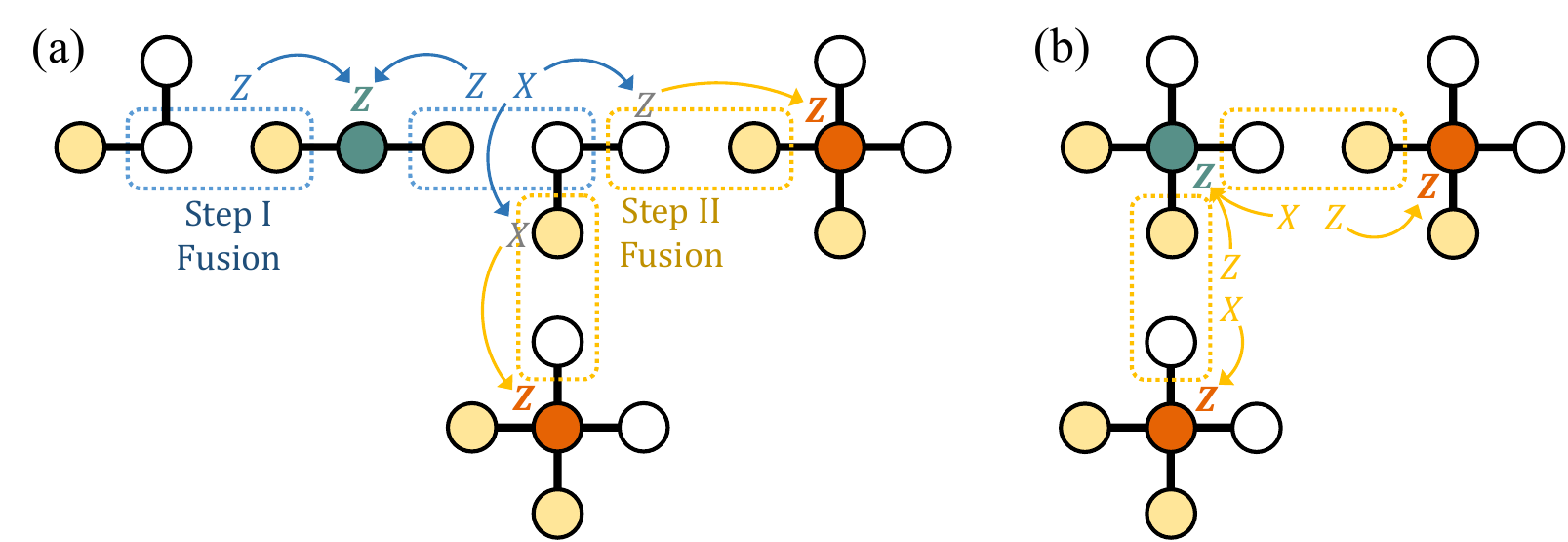}
        \caption{Propagation of errors from (a) Step I fusion and (b) Step II fusion. We only track error propagation resulting in $Z$ errors on data qubits. Note that qubits with Hadamard gates are filled with yellow.}
        \label{fig:Propagation}
    \end{figure}

In our simulation, we track only $Z$ errors of data qubits because we only simulate errors on primal lattices and $X$ errors are not detected by Pauli $X$ measurements. We show error propagation during RHG lattice construction in Fig.~\ref{fig:Propagation}. In Step 1, two fusion operations are performed respectively on the side qubits of $\ket{C_3^\mathrm{c}}$. A $Z$ error in the fusion causes $Z$ error on the data qubit and an $X$ error causes errors on the side qubits of $\ket{C_5^\ast}$. Errors on the side qubits further propagate to the adjacent data qubits causing $Z$ errors in the next step. In Step 2, a $Z$($X$) error in the fusion propagates causing $Z$ error on the date qubit in the direction with(without) the Hadamard gate, following to the propagation rule.

\section{\label{sec:Simulation}Simulation details}

\subsection{RHG lattice}

We here describe the simulation details of RHG lattice. We perform a Monte Carlo simulation to find the logical error rate and the sequence of each trial is as follows:
\begin{enumerate}
    \item We assign the $Z$ error rate to each primal qubit $j$ on the lattice and denote it by $q_{Z,j}$. The error rate includes the propagated one from the fusion operation as well as the $Z$ error due to photon losses.
    \item The errors on qubits are randomly sampled according to the error rate. We determine the value of the syndrome measurement on each cell using the errors of the qubits located on faces of the cell.
    \item The syndromes are decoded to find the error pattern using the weighted minimum-weight perfect matching \cite{Fowler2012} in the PyMatching package \cite{Higgott2021}, where the weight for each qubit is given by $ \log\left[(1-q_{Z,j})/q_{Z,j}\right] $.
    \item Once we remove the detected errors, the remaining error chains connect two opposite $x$ boundaries which are primal. If the number of remaining error chains is odd, we identify a logical error.
\end{enumerate}

By repeating the trial $3\times10^5$ times, we decide the logical error rate $p_L$. The logical error rates are calculated with code distances $ d = 3, 5 $ while varying the loss rate $\eta$. The loss threshold $\eta_\textrm{th}$ is obtained by finding the largest $\eta$ such that $p_L$ decreases with $d$.

\subsection{Steane code}

For circuit-based quantum computation, we simulate a telecorrection circuit of seven-qubit Steane code with several levels of concatenation. The circuit is composed of the preparation of $\ket{+_L}$, $CZ$ gate, $H$ gate, and $X$-basis measurement. In our simulation, we postselect the successful gate teleportation without error to prevent errors from propagating further. Therefore, there remains only unlocatable errors: one is due to photon loss and the other is from the $Z$ error of fusion operation. In the $H$ gate teleportation, a $Z$ error of fusion operation induces an $X$ error on the output qubit because the Hadamard gate is applied on the output qubit. In the $CZ$ gate teleportation, a $Z$ error of each fusion operation induces a $Z$ error on one of the output qubit. When qubits are stored in quantum memory, photon loss induces Pauli $Z$ error.

    \begin{figure}[t]
        \centering \includegraphics[clip=true, width=0.8\columnwidth]{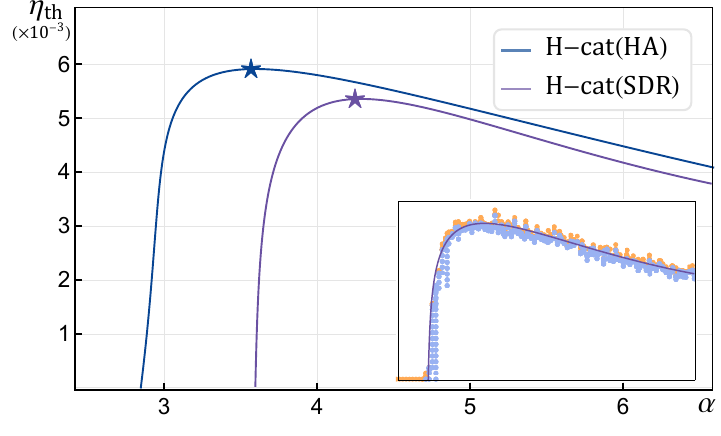}
        \caption{Threshold curve for Steane code simulation with HA and SDR scheme. Points marked as stars represent the highest loss threshold. In the inset, we present the way of determining the threshold curve. From a Monte Carlo simulation with parameters $\eta$ and $\alpha$, we decide if parameters are accepted(rejected) for fault-tolerant computation, which are marked as blue(orange). The threshold curve is determined to be the boundary between two regions.}
        \label{fig:Steane}
    \end{figure}
We find the logical error rate of locatable and unlocatable errors using the Monte Carlo method. In the first level of concatenation, we use the error rates of H-cat qubits with the loss rate $\eta$ and the amplitude $\alpha$. The resulting error rates are used for the next level of concatenation. If the error rates tend to decrease for further levels of concatenation, we assure that error correction works in a fault-tolerant way with the given value of $\eta$ and $\alpha$. We repeat the simulation while varying $\eta$ and $\alpha$ and obtain the loss threshold curve as shown in Fig.~\ref{fig:Steane}.

It is shown that HA scheme achieves a slightly higher threshold than SDR scheme. However, in the circuit-based error correction, HA scheme is not efficient because we postselect the successful gate teleportation without error and the located error rate of HA scheme is of constant order as shown in Fig.~\ref{fig:ErrBellCat}. Rather SDR scheme is more appropriate due to the near-deterministic success probability of fusion operation.

\end{document}